\documentclass[11pt, a4paper, fleqn]{article}
\usepackage{amsmath}
\usepackage{geometry}
\usepackage{amssymb}  

\numberwithin{equation}{section}
                                   
\begin{document}

\setlength{\abovedisplayskip}{1 mm}
\setlength{\belowdisplayskip}{1 mm}

\title {On some properties of the electromagnetic field\\and its interaction with a charged particle}
\author{Dimitar Simeonov}
\maketitle

\begin{abstract}
{A procedure for solving the Maxwell equations in vacuum, under the additional requirement that both scalar invariants are equal to zero, is presented. Such a field is usually called a null electromagnetic field. Based on the complex Euler potentials that appear as arbitrary functions in the general solution, a vector potential for the null electromagnetic field is defined. This potential is called natural vector potential of the null electromagnetic field. An attempt is made to make the most of knowing the general solution. The properties of the field and the potential are studied without fixing a specific family of solutions. A equality, which is similar to the Dirac gauge condition, is found to be true for both null field and Lienard-Wiechert field. It turns out that the natural potential is a substantially complex vector, which is equivalent to two real potentials. A modification of the coupling term in the Dirac equation is proposed,  that makes the equation work with both real potentials. A solution, that corresponds to the Volkov's solution for a Dirac particle in a linearly polarized plane electromagnetic wave, is found. The solution found is directly compared to Volkov's solution under the same conditions.}
\end{abstract}

\section{Complex notation}
To denote the electromagnetic field we use the complex vector
\begin{equation} \label{ComplexField}
\textbf{F} = \textbf{E} + ic\textbf{B}
\end{equation}
where \textbf{E} and \textbf{B} are the real vectors of the electric and magnetic field and $c$ is the speed of light. This vector has been used by Riemann, Silberstein, Bateman, Majorana, Kramers and many others for over a century. We refer to it simply as electromagnetic field.
Maxwell equations for this complex vector in SI units are
\begin{equation} \label{Maxw1}
\boldsymbol{\nabla} \textbf{.F} = \rho/\epsilon_0 \hspace{1 cm}
\partial_t \textbf{F} + i\boldsymbol{\nabla} \times \textbf{F} = -\mu_0c \textbf{I}
\end{equation}
where $\rho$ and $\textbf{I}$ are the densities of the electric charge and current and $\epsilon_0$ and $\mu_0$ are dielectric and magnetic constant of vacuum.
In four-vector form
\begin{equation}
\partial{_\mu} F^{\mu\nu} = \mu_0c I^\nu \hspace{20 mm}
\partial^{\mu} = (\partial_t, -\boldsymbol{\nabla}), \hspace{6 mm}
\partial_t = \partial_{ct}, \hspace{6 mm}
I^{\mu} = (\rho c, \textbf{I})
\end{equation}
The antisymmetric complex field tensor is
\begin{equation}
F^{\mu\nu} =
\begin{vmatrix}
0 & -F_x & -F_y & -F_z\\
F_x & 0 & iF_z & -iF_y\\
F_y & -iF_z & 0 & iF_x\\
F_z & iF_y & -iF_x & 0
\end{vmatrix}
\end{equation}
The two real scalar invariants are written as a single complex value
\begin{equation}
\textbf{F}^2 =  \textbf{E}^2 - c^2\textbf{B}^2 + 2ic\textbf{E.B}
\end{equation}
The electromagnetic energy density and the Poynting vector (omitting the factor $\epsilon_0/2$) are
\begin{equation}
\textbf{E}^2 + c^2\textbf{B}^2 = \textbf{F.F}^* \hspace{1 cm}
2\textbf{E} \times c\textbf{B} = i\textbf{F} \times \textbf{F}^*
\end{equation}

\section{Spinor representation of the electromagnetic field}
The electromagnetic field \eqref{ComplexField} can always be written as
\begin{equation} \label{SpRep}
\textbf{F} = \phi^{\dagger}\boldsymbol{\sigma}\psi =
(\phi_1^*\psi_2 + \phi_2^*\psi_1, -i\phi_1^*\psi_2 + i\phi_2^*\psi_1, \phi_1^*\psi_1 - \phi_2^*\psi_2)
\end{equation}
where $\phi$ and $\psi$ are one left-handed and one right-handed two-component Weyl spinor and $\boldsymbol{\sigma}$ are the Pauli matrices. This representation of the electromagnetic field was suggested and discussed in detail by Campolattaro$^{[1]}$ who used Dirac spinors in his study. Here we use Weyl spinors which allows us to write very simple vector equations. Using these spinors, we can construct some other tensors that are not at our disposition if we work with the electromagnetic field \textbf{F} only.

We define some tensors and identities here and refer to them when needed in the rest of the text. Besides the complex scalar
\begin{equation}
s = \phi^{\dagger}\psi = \phi_1^*\psi_1 + \phi_2^*\psi_2
\hspace{20 mm} s^* = \psi^{\dagger}\phi
\end{equation}
one has two chiral currents
\begin{equation}
J_\textsc{R} = (J_\textsc{R0}, \textbf{J}_\textsc{R}) = (\psi^{\dagger}\psi, \psi^{\dagger}\boldsymbol{\sigma}\psi)
\end{equation}
\begin{equation*}
\hspace{5 mm} = (\psi_1^*\psi_1 + \psi_2^*\psi_2, \psi_1^*\psi_2 + \psi_2^*\psi_1, 
-i\psi_1^*\psi_2 + i\psi_2^*\psi_1, \psi_1^*\psi_1 - \psi_2^*\psi_2)
\end{equation*}
\begin{equation*}
J_\textsc{L} = (J_\textsc{L0}, \textbf{J}_\textsc{L}) = (\phi^{\dagger}\phi, -\phi^{\dagger}\boldsymbol{\sigma}\phi)
\end{equation*}
\begin{equation*}
\hspace{5 mm} = (\phi_1^*\phi_1 + \phi_2^*\phi_2, -\phi_1^*\phi_2 - \phi_2^*\phi_1, 
i\phi_1^*\phi_2 - i\phi_2^*\phi_1, -\phi_1^*\phi_1 + \phi_2^*\phi_2)
\end{equation*}

A number of purely formal identities connect the tensors that can be constructed by two spinors. These identities are simply a consequence of the definitions of the tensors, and by themselves do not contain any physics. For example, one has
\begin{equation}
\textbf{F}^2 = s^2 \hspace{7 mm}
J_\textsc{R0}^2 - \textbf{J}_\textsc{R}^2 = 0 \hspace{7 mm} 
J_\textsc{L0}^2 - \textbf{J}_\textsc{L}^2 = 0 \hspace{7 mm} 
J_\textsc{R0} J_\textsc{L0} - \textbf{J}_\textsc{R}\textbf{.}\textbf{J}_\textsc{L} = 2ss^*
\end{equation}
and also
\begin{equation}
\textbf{F.J}_\textsc{R} = +sJ_\textsc{R0} \hspace{10 mm}
\textbf{F}J_\textsc{R0} + i\textbf{F} \times \textbf{J}_\textsc{R} = +s\textbf{J}_\textsc{R}
\end{equation}
\begin{equation*}
\textbf{F.J}_\textsc{L} = -sJ_\textsc{L0} \hspace{11 mm}
\textbf{F}J_\textsc{L0} + i\textbf{F} \times \textbf{J}_\textsc{L} = -s\textbf{J}_\textsc{R}
\end{equation*}
which in four-vector form is
\begin{equation*}
F^{\mu\nu}J_{\textsc{R}\nu} = +sJ_\textsc{R}^{\mu}\hspace{8 mm}
F^{\mu\nu}J_{\textsc{L}\nu} = -sJ_\textsc{L}^{\mu}
\end{equation*}
That is, the chiral currents $J_\textsc{R}$ and $J_\textsc{L}$ are the eigenvectors of the complex field tensor with eigenvalues $+s$ and $-s$, respectively. The spinor representation of the field actually solves the eigenvalue problem for the complex field tensor.
The electromagnetic energy-momentum tensor (omitting the factor $\epsilon_0/2$) is
\begin{equation*}
T^{\mu\nu} = J_\textsc{R}^{\mu}J_\textsc{L}^{\nu} + J_\textsc{R}^{\nu}J_\textsc{L}^{\mu} - ss^*g^{\mu\nu}
\hspace{20 mm} g = diag(1, -1, -1, -1)
\end{equation*}
One may also need the currents
\begin{equation} \label{DiracCurrent}
J_\textsc{D} = J_\textsc{R} + J_\textsc{L} \hspace{20 mm} 
J_\textsc{A} = J_\textsc{R} - J_\textsc{L}
\end{equation}
Using this currents the electromagnetic energy-momentum tensor is written as
\begin{equation} \label{EMTbyDiracCurrent}
2T^{\mu\nu} = 
J_\textsc{D}^{\mu}J_\textsc{D}^{\nu} - J_\textsc{A}^{\nu}J_\textsc{A}^{\mu} 
- 2ss^*g^{\mu\nu}
\end{equation}

If $\phi$ and $\psi$ were spinors representing a wave function of a Dirac particle, the currents \eqref{DiracCurrent} would be the probability current and the axial current in Dirac theory respectively. In this study we use "Dirac current" and "axial current" to refer to the currents \eqref{DiracCurrent}, whether an electromagnetic field expressed by spinors, or a wave function of a Dirac particle, is in consideration.

\section{General solution to the null electromagnetic field problem}
We will find the general solution of the Maxwell equations in vacuum
\begin{equation} \label{Maxw2}
\boldsymbol{\nabla} \textbf{.F} = 0 \hspace{1 cm}
\partial_t \textbf{F} + i\boldsymbol{\nabla} \times \textbf{F} = 0
\end{equation}
requiring both scalar invariants to be zero
\begin{equation}
\textbf{F}^2 = 0
\end{equation}
A field that meets these conditions is usually called null electromagnetic field. It is generally accepted that such a field represents "pure radiation".

To reduce a little bit writing in what follows we make a change of variables
\begin{equation}
x_1 = x - iy \hspace{10 mm} z_1 = z - ct
\end{equation}
\begin{equation*}
x_2 = x + iy \hspace{10 mm} z_2 = z + ct
\end{equation*}
Maxwell equations \eqref{Maxw2} are
\begin{equation} \label{Maxw3}
\partial _{x_1}F_z - \partial _{z_1}(F_x + iF_y) = 0 \hspace{10 mm}
\partial _{x_2}(F_x + iF_y) + \partial _{z_2}F_z = 0 
\end{equation}
\begin{equation*}
\partial _{x_1}(F_x - iF_y) + \partial _{z_1}F_z = 0 \hspace{10 mm}
\partial _{x_2}F_z - \partial _{z_2}(F_x - iF_y) = 0
\end{equation*}

The spinor representation of the electromagnetic field \eqref{SpRep} introduces four new unknown complex functions. However, the real number of unknowns is only two, given that the square of the field is zero. This allows us to impose an additional condition connecting the two spinors $\phi$ and $\psi$.
\begin{equation} \label{NulCond}
\phi = -i\sigma_y \psi^* \hspace{15 mm}
(\phi_1 = -\psi_2^*, \phi_2 = \psi_1^*)
\end{equation}
This is formally identical to the well known Majorana condition. However, the Majorana condition is used in a completely different context, so we cannot simply call the equation \eqref{NulCond} Majorana condition. Here we call it null field condition. The null field condition reduces the number of unknown functions to two. The square of the field is zero since $\phi^{\dagger}\psi$ is zero when \eqref{NulCond} is satisfied. The null field condition is Lorentz invariant and will be valid in every inertial reference frame.

Under the null field condition the electromagnetic field \eqref{SpRep} is
\begin{equation} \label{CartanSpinor}
\textbf{F} = \psi^T i {\sigma}_y \boldsymbol{\sigma}\psi =
(\psi_1^2 - \psi_2^2, i\psi_1^2 + i\psi_2^2, -2\psi_1\psi_2)
\end{equation}
Maxwell equations \eqref{Maxw3} for $\psi$ are
\begin{equation} \label{Maxw4}
\partial _{x_1}\psi_1^2 - \partial _{z_1}\psi_1\psi_2 = 0 \hspace{10 mm}
\partial _{x_2}\psi_1\psi_2 + \partial _{z_2}\psi_1^2 = 0
\end{equation}
\begin{equation*}
\partial _{x_1}\psi_1\psi_2 - \partial _{z_1}\psi_2^2 = 0 \hspace{10 mm}
\partial _{x_2}\psi_2^2 + \partial _{z_2}\psi_1\psi_2 = 0
\end{equation*}
To effectively solve this system it is convenient to put
\begin{equation} \label{sub1}
\psi_1^2 = u \hspace{20 mm} \psi_2/\psi_1 = v
\end{equation}
Inserting \eqref{sub1} into \eqref{Maxw4}, we obtain the following equations for the new unknown functions $u$ and $v$.
\begin{equation} \label{Maxw5}
(\partial_{x_1} - v \partial_{z_1}) v = 0 \hspace{8 mm}
(\partial_{x_1} - v \partial_{z_1}) u - u\partial_{z_1} v = 0
\end{equation}
\begin{equation*}
(\partial_{z_2} + v \partial_{x_2}) v = 0 \hspace{8 mm}
(\partial_{z_2} + v \partial_{x_2}) u + u\partial_{x_2} v = 0
\end{equation*}
The two equations containing only the function v are of a familiar type (inviscid Burgers equation) and their solution is known. It is
\begin{equation*}
v = f(z_1 + v x_1, x_2 - v z_2)
\end{equation*}
where $f(\xi, \eta)$ is an arbitrary complex function of two variables.
We notice, however, that instead of \eqref{sub1} we can use the substitution
\begin{equation} \label{sub2}
\psi_2^2 = u \hspace{20 mm} \psi_1/\psi_2 = v
\end{equation}
Doing this and following the steps above we get a different solution for the function $v$
\begin{equation*}
v = f(z_2 - v x_2, x_1 + v z_1)
\end{equation*}
This observation suggests that we can try to find the general solution of the system \eqref{Maxw5} by the following change of variables
\begin{equation}
\xi_1 = z_1 + v x_1 \hspace{10 mm} \eta_1 = x_2 - v z_2
\end{equation}
\begin{equation*}
\xi_2 = z_2 - v x_2 \hspace{10 mm} \eta_2 = x_1 + v z_1
\end{equation*}
The corresponding derivatives are
\begin{align*}
&\partial_{x_1} = (v + x_1\partial_{x_1}v)\partial_{\xi_1} - z_2 (\partial_{x_1}v)\partial_{\eta_1} - x_2(\partial_{x_1}v)\partial_{\xi_2} + (1 + z_1\partial_{x_1}v)\partial_{\eta_2}& \\
&\partial_{z_1} = (1 + x_1\partial_{z_1}v)\partial_{\xi_1} - z_2 (\partial_{z_1}v)\partial_{\eta_1} - x_2(\partial_{z_1}v)\partial_{\xi_2} + (v + z_1\partial_{z_1}v)\partial_{\eta_2}& \\
&\partial_{x_2} = x_1(\partial_{x_2}v)\partial_{\xi_1} + (1 - z_2 \partial_{x_2}v)\partial_{\eta_1} - (v + x_2 \partial_{x_2}v)\partial_{\xi_2} + z_1(\partial_{x_2}v)\partial_{\eta_2}& \\
&\partial_{z_2} = x_1(\partial_{z_2}v)\partial_{\xi_1} - (v + z_2 \partial_{z_2}v)\partial_{\eta_1} + (1 - x_2 \partial_{z_2}v)\partial_{\xi_2} + z_1(\partial_{z_2}v)\partial_{\eta_2}& 
\end{align*}
In particular the derivatives of the function $v$ are
\begin{align*}
\partial_{x_1}v & = ( v\partial_{\xi_1}v + \partial_{\eta_2}v )/R &
\partial_{x_2}v & = ( -v\partial_{\xi_2}v + \partial_{\eta_1}v )/R \\
\partial_{z_1}v & = ( \partial_{\xi_1}v + v\partial_{\eta_2}v )/R &
\partial_{z_2}v & = ( \partial_{\xi_2}v - v\partial_{\eta_1}v )/R
\end{align*}
where we have put for brevity
\begin{equation*}
R = 1 - x_1\partial_{\xi_1}v + z_2\partial_{\eta_1}v + x_2\partial_{\xi_2}v - z_1\partial_{\eta_2}v
\end{equation*}
Now we can calculate the differential operators in \eqref{Maxw5}
\begin{align*}
\partial_{x_1} - v\partial_{z_1} & = (1 - v^2)\partial_{\eta_2} + (\partial_{x_1}v - v\partial_{z_1}v)(x_1\partial_{\xi_1} - z_2\partial_{\eta_1} - x_2\partial_{\xi_2} + z_1\partial_{\eta_2})  \\
\partial_{z_2} + v\partial_{x_2} & = -(1 + v^2)\partial_{\xi_2} + (\partial_{z_2}v + v\partial_{x_2}v)(x_1\partial_{\xi_1} - z_2\partial_{\eta_1} - x_2\partial_{\xi_2} + z_1\partial_{\eta_2})
\end{align*}
and the expressions
\begin{align*}
(\partial_{x_1} - v\partial_{z_1})v & = (1 - v^2)(\partial_{\eta_2}v)/R \\
(\partial_{z_2} + v\partial_{x_2})v & = (1 - v^2)(\partial_{\xi_2}v)/R
\end{align*}
We finally get the system
\begin{equation*}
\partial_{\xi_2}v = 0
\end{equation*}
\begin{equation*}
\partial_{\eta_2}v = 0
\end{equation*}
\begin{equation*}
\partial_{\xi_2}u = -u(\partial_{\eta_1}v)/[\ 1 - v^2 - (\eta_2 - v\xi_1)\partial_{\xi_1}v + (\xi_2 + v\eta_1)\partial_{\eta_1}v\ ]
\end{equation*}
\begin{equation*}
\partial_{\eta_2}u = u(\partial_{\xi_1}v)/[\ 1 - v^2 - (\eta_2 - v\xi_1)\partial_{\xi_1}v + (\xi_2 + v\eta_1)\partial_{\eta_1}v\ ]
\end{equation*}
When writing the second pair of equations we have taken into account that according to the first pair of equations, $v$ does not depend on $\xi_2$ and $\eta_2$.

All four equations of this system are easily integrated. The first pair of equations gives the obvious result $v = f(\xi_1, \eta_1)$ where $f$ is an arbitrary complex function of two variables. The solution to the third equation is
\begin{equation*}
u = g_1(\xi_1, \eta_1, \eta_2)/[\ 1 - v^2 - (\eta_2 - v\xi_1)\partial_{\xi_1}v 
+ (\xi_2 + v\eta_1)\partial_{\eta_1}v\ ]
\end{equation*}
and the solution to the fourth equation is
\begin{equation*}
u = g_2(\xi_1, \eta_1, \xi_2)/[\ 1 - v^2 - (\eta_2 - v\xi_1)\partial_{\xi_1}v 
+ (\xi_2 + v\eta_1)\partial_{\eta_1}v\ ]
\end{equation*}
where $g_1$ and $g_2$ are arbitrary complex functions. The latest two expressions represent the same function and must be identical so that
\begin{equation*}
g_1(\xi_1, \eta_1, \eta_2) = g_2(\xi_1, \eta_1, \xi_2) = g(\xi_1, \eta_1)
\end{equation*}

\vspace{2 mm}

Let us summarize the final result and return to real coordinates $(\textbf{r}, t)$.

1. The function $v(\textbf{r}, t)$ is implicitly defined by the equation
\begin{equation} \label{eqforv}
v(\textbf{r}, t) = f(\xi, \eta)
\end{equation}
where
\begin{equation} \label{xieta}
\xi(\textbf{r}, t) = z - ct + (x - iy) v(\textbf{r}, t)
\end{equation}
\begin{equation*}
\eta(\textbf{r}, t) = x + iy - (z + ct) v(\textbf{r}, t)
\end{equation*}
and $f(\xi, \eta)$ is an arbitrary complex function of two variables.

2. The function $u(\textbf{r}, t)$ is defined as
\begin{equation}
u(\textbf{r}, t) = g(\xi, \eta) / R
\end{equation}
where
\begin{equation}
R(\textbf{r}, t) = 1 - (x - iy) f_\xi + (z + ct) f_\eta
\end{equation}
and $g(\xi, \eta)$ is second arbitrary complex function of two variables and $f_\xi$ and $f_\eta$ are the partial derivatives of $f$ with respect to the corresponding arguments.

3. According to \eqref{CartanSpinor} and \eqref{sub1}, the electromagnetic field is
\begin{equation}
\textbf{F} = u [\ (1 - v^2), i(1 + v^2), -2v\ ]
\end{equation}
The latter can be written as
\begin{equation} \label{FbyF0}
\textbf{F} = g(\xi, \eta) \textbf{F}_0
\end{equation}
where
\begin{equation} \label{PrimaryField}
\textbf{F}_0 = \frac{1}{R} [\ (1 - v^2), i(1 + v^2), -2v\ ]
\end{equation}

Each choice of an arbitrary complex function of two variables $f(\xi, \eta)$ defines a family of solutions. This function could be called a generating function of the family or simply a family generator. The equation \eqref{PrimaryField} defines a primary field $\textbf{F}_0$ for each family. All other members of the family can be derived by multiplying $\textbf{F}_0$ by an arbitrary amplitude function $g(\xi, \eta)$. The superposition rule applies only to members of one family. A sum of solutions from different families produces a field with non zero square. Note that the primary field $\textbf{F}_0$ is dimensionless. The dimension is incorporated in the amplitude function $g(\xi, \eta)$.

For the energy density and the Poynting vector (omitting the factor $\epsilon_0/2$) one has
\begin{equation*}
\textbf{F.F}^* = 2uu^*(1 + vv^*)^2
\end{equation*}
\begin{equation*}
i\textbf{F} \times \textbf{F}^* = 2uu^*(1 + vv^*)[\ v + v^*, -i(v - v^*), 1 - vv^*\ ]
\end{equation*}
We can define a real unit vector $\textbf{n}$ that determines the direction of the Poynting vector at each point in space.
\begin{equation} \label{upv}
\textbf{n} = [\ v + v^*, -i(v - v^*), 1 - vv^*\ ]/(1 + vv^*)
\end{equation}
\begin{equation*}
\textbf{n}^2 = 1, \hspace{6 mm} i\textbf{F} \times \textbf{F}^* = (\textbf{F.F}^*)\textbf{n}, \hspace{6 mm}\textbf{F} - i\textbf{n} \times \textbf{F} = 0, \hspace{6 mm}\textbf{n.F} = 0
\end{equation*}
This vector depends only on the choice of a family generator $f$ and is the same for all members of one family.
It can be viewed (up to a multiplier) as the space part of a true four-vector.
According to \eqref{sub1} the components of the spinor $\psi$ are
\begin{equation}
\psi_1 = \sqrt{u} \hspace{10 mm}
\psi_2 = v\sqrt{u}
\end{equation}
Under the null field condition \eqref{NulCond} one has 
$J_\textsc{L} = J_\textsc{R}$, $J_\textsc{A} = 0$ and $J_\textsc{D} = 2J_\textsc{R}$, so there is only one current available.
The Dirac current is
\begin{equation*}
J_\textsc{D} = 2|u|[\ 1 + vv^*, v + v^*, -i(v - v^*), 1 - vv^*\ ]
= 2|u|(1 + vv^*)(1, \textbf{n})
\end{equation*}
The electromagnetic energy-momentum tensor \eqref{EMTbyDiracCurrent} with $s = 0$ and $J_\textsc{A} = 0$ is
\begin{equation*}
T^{\mu\nu} = \frac{1}{2}J_\textsc{D}^{\mu}J_\textsc{D}^{\nu}
\end{equation*}
A system with this type of energy-momentum tensor could be viewed as a "dust".
It is natural to ask the question whether Dirac current $J_\textsc{D}$ can be physically interpreted in some way. Without going into detail, we note that in general $\partial_t J_\textsc{D0} + \boldsymbol{\nabla} \textbf{.J}_\textsc{D} \neq 0$ for a null electromagnetic field. So this current can not be interpreted as a density and flux of a locally conserved physical quantity.

If one use the substitution \eqref{sub2} instead of \eqref{sub1} one gets the general solution in another form.
\begin{equation} \label{form2}
v(\textbf{r}, t) = f(\xi, \eta)
\end{equation}
\begin{equation*}
\xi(\textbf{r}, t) = z + ct - (x + iy) v(\textbf{r}, t) \hspace{8 mm}
\eta(\textbf{r}, t) = x - iy + (z - ct) v(\textbf{r}, t)
\end{equation*}
\begin{equation*}
u(\textbf{r}, t) = g(\xi, \eta) / [\ 1 + (x + iy) f_\xi - (z - ct) f_\eta\ ] \\
\end{equation*}
\begin{equation*}
\textbf{F} = u [\ (v^2 - 1), i(v^2 + 1), -2v\ ]
\hspace{8 mm}
\textbf{n} = [\ v + v^*, i(v - v^*), vv^* - 1\ ]/(1 + vv^*)
\end{equation*}

The simplest possible choice for the family generator $f$ is $f(\xi, \eta) = const$. In this case one has $\xi = z - ct + (x - iy)v$, $\eta = x + iy - (z + ct)v$
with $v = const$. The unit Poynting vector $\textbf{n}$ \eqref{upv} is constant everywhere, so this solution is a plane wave. Depending on the value of the complex constant $v$, which is equivalent to two real parameters, we will have waves with different directions of propagation in space. All directions are possible except for the negative direction of the $z$ axis. It is because there is no complex value $v$ that solves the equation $n_z = (1 - vv^*) / (1 + vv^*) = -1$.
Instead of the complex constant $v$, we can use two real angles $\vartheta$ and $\varphi$ that describe the direction of the unit vector $\textbf{n}$.
\begin{equation*}
\textbf{n} = [\ sin(\vartheta) cos(\varphi), sin(\vartheta) sin(\varphi), cos(\vartheta)\ ]
\end{equation*}
Then $v = \tan(\vartheta / 2) e^{i\varphi}$ which is not defined for $\vartheta = \pi$.
The special case $\vartheta = \pi$ is covered by the second form of the solution
\eqref{form2}. For this form we have
$v = \cot(\vartheta / 2) e^{-i\varphi}$
which is defined for $\vartheta = \pi$ but is not defined for $\vartheta = 0$. All directions are possible except for the positive direction of the z axis.
The general expression for a plane wave traveling along z axis is
\begin{equation*}
\textbf{F} = g(z - ct, x + iy) (1, i, 0)
\end{equation*}
for the positive direction of propagation and
\begin{equation*}
\textbf{F} = g(z + ct, x - iy) (1, -i, 0)
\end{equation*}
for the negative direction of propagation.

Let us look at one more classical example of a null field. If we choose the family generator as
$f(\xi, \eta) = \xi / \eta$
the equation \eqref{eqforv} becomes a simple quadratic equation for $v$
\begin{equation*}
(x - iy) v^2 + 2z v - (x + iy) = 0
\end{equation*}
One of the two roots of this equation is $v = (x + iy)/(r + z)$.
For the unit Poynting vector one has $\textbf{n} = \textbf{r} / r$.
Therefore, this solution is an outgoing spherical wave. The general expression for an outgoing null spherical wave is
\begin{equation*}
\textbf{F} = \frac{1}{r} g(r - ct, v) [\ 1 - v^2, i(1 + v^2), -2v\ ]
\end{equation*}
with arbitrary function $g$ and
$ v = \tan(\vartheta / 2) e^{i\varphi} $
in spherical coordinates. The second root of the quadratic equation gives an incoming null spherical wave.
\vspace{2 mm}

We will not consider other specific examples of null fields here. The simplest solutions are already known and there is no point in repeating them. Our intention is to derive as many general null field properties as possible without focusing on a particular family. We want to make the most of knowing the general solution.

The arbitrary functions used in constructing the general solution depend on the coordinates and time only through the variables $\xi$ and $\eta$. The derivatives with respect to the coordinates and time of such functions are not independent. They are connected by some algebraic relations. These algebraic relations determine most properties of the null field.
From the definition of $\xi$ and $\eta$ \eqref{xieta} for the derivatives of each function $\Phi(\xi, \eta)$ we obtain
\begin{equation} \label{Derivatives1}
\partial_t \Phi = - [\ 1 - (x - iy)\partial_t v\ ]\Phi_{\xi} 
- [\ v + (z + ct)\partial_t v\ ]\Phi_{\eta}
\end{equation}
\begin{equation*}
\partial_x \Phi = [\ v + (x - iy)\partial_x v\ ]\Phi_{\xi} 
+ [\ 1 - (z + ct)\partial_x v\ ]\Phi_{\eta}
\end{equation*}
\begin{equation*}
\partial_y \Phi = 
[\ -iv + (x - iy)\partial_y v\ ]\Phi_{\xi} 
+ [\ i - (z + ct)\partial_y v\ ]\Phi_{\eta}
\end{equation*}
\begin{equation*}
\partial_z \Phi = 
[\ 1 + (x - iy)\partial_z v\ ]\Phi_{\xi} 
- [\ v + (z + ct)\partial_z v\ ]\Phi_{\eta}
\end{equation*}
where $\Phi_{\xi}$ and $\Phi_{\eta}$ are the partial derivatives of $\Phi$ with respect to the corresponding arguments. Putting $\Phi = v(\textbf{r}, t) = f(\xi, \eta)$ and solving each of the equations with respect to corresponding derivative we obtain
\begin{equation} \label{Derivatives2}
\begin{split}
\partial_t v & = -(f_{\xi} + f f_{\eta})/R \\
\partial_z v & = (f_{\xi} - f f_{\eta})/R
\end{split}
\qquad\qquad
\begin{split}
\partial_x v & = (f f_{\xi} + f_{\eta})/R \\
\partial_y v & = i(-f f_{\xi} + f_{\eta})/R
\end{split}
\end{equation}
It follows that
\begin{align*}
(\partial_t + \partial_z) v + v(\partial_x - i \partial_y) v = 0 \\
(\partial_x + i \partial_y) v + v(\partial_t - \partial_z) v = 0
\end{align*}
\begin{equation*}
(\partial_t v)^2 - (\boldsymbol{\nabla} v)^2 = 0
\end{equation*}
The derivatives of $\xi$ and $\eta$ are
\begin{equation} \label{Derivatives3}
\begin{split}
\partial_t\xi & = -1 + (x - iy) \partial_t v  \\
\partial_x\xi & = v + (x - iy) \partial_x v   \\
\partial_y\xi & = -iv + (x - iy) \partial_y   \\
\partial_z\xi & = 1 + (x - iy) \partial_z v
\end{split}
\qquad\qquad
\begin{split}
\partial_t\eta & = -v - (z + ct) \partial_t v  \\
\partial_x\eta & = 1 - (z + ct) \partial_x v   \\
\partial_y\eta & = i - (z + ct) \partial_y v   \\
\partial_z\eta & = -v - (z + ct) \partial_z v
\end{split}
\end{equation}

Using \eqref{Derivatives1}, \eqref{Derivatives2} and \eqref{Derivatives3}, by simple purely algebraic calculations, one can verify the following equalities
\begin{equation}
(\partial_t \xi)^2 - (\boldsymbol{\nabla} \xi)^2 = 0 \hspace{10 mm}
(\partial_t \eta)^2 - (\boldsymbol{\nabla} \eta)^2 = 0
\end{equation}
\begin{equation*}
(\partial_t \xi)(\partial_t \eta) - (\boldsymbol{\nabla} \xi)\textbf{.}(\boldsymbol{\nabla} \eta) 
= 0
\end{equation*}
\begin{equation*}
(\partial_t \eta)(\boldsymbol{\nabla} \xi) - (\partial_t \xi)(\boldsymbol{\nabla} \eta)
= [\ (1 - v^2), i(1 + v^2), -2v\ ] / R
\end{equation*}
\begin{equation*}
i(\boldsymbol{\nabla} \xi) \times (\boldsymbol{\nabla} \eta)
= [\ (1 - v^2), i(1 + v^2), -2v\ ] / R
\end{equation*}
The right hand sides in the latest two equations coincide with the primary field for the family $\textbf{F}_0$. So
\begin{equation}
\textbf{F}_0 = (\partial_t \eta)(\boldsymbol{\nabla} \xi) - (\partial_t \xi)(\boldsymbol{\nabla} \eta)
= i(\boldsymbol{\nabla} \xi) \times (\boldsymbol{\nabla} \eta)
\end{equation}
It also follows that for each function $\Phi(\xi, \eta)$
\begin{align} \label{DerivativesRelation}
(\partial_t + \partial_z) \Phi + v(\partial_x - i \partial_y) \Phi = 0
\end{align}
\begin{align*}
(\partial_x + i \partial_y) \Phi + v(\partial_t - \partial_z) \Phi = 0
\end{align*}
\begin{equation*}
(\partial_t \Phi)^2 - (\boldsymbol{\nabla} \Phi)^2 = 0
\end{equation*}
And for each pair of functions $\Phi_1(\xi, \eta)$ and $\Phi_2(\xi, \eta)$
\begin{equation} \label{NoName301}
(\partial_t\Phi_2)(\boldsymbol{\nabla}\Phi_1) - (\partial_t\Phi_1)(\boldsymbol{\nabla}\Phi_2) =
[\ (\partial_\xi\Phi_1)(\partial_\eta\Phi_2) - (\partial_\eta\Phi_1)(\partial_\xi\Phi_2)\ ] \textbf{F}_0
\end{equation}
\begin{equation} \label{NoName302}
i(\boldsymbol{\nabla}\Phi_1) \times (\boldsymbol{\nabla}\Phi_2) =
[\ (\partial_\xi\Phi_1)(\partial_\eta\Phi_2) - (\partial_\eta\Phi_1)(\partial_\xi\Phi_2)\ ] \textbf{F}_0
\end{equation}
Comparing \eqref{NoName301} and \eqref{NoName302} with \eqref{FbyF0}, one can see that the general solution can be represented in the following two equivalent forms
\begin{equation} \label{BatemanField1}
\textbf{F} = (\partial_t\Phi_2)(\boldsymbol{\nabla}\Phi_1) - (\partial_t\Phi_1)(\boldsymbol{\nabla}\Phi_2)
\end{equation}
\begin{equation} \label{BatemanField2}
\textbf{F} = i(\boldsymbol{\nabla}\Phi_1) \times (\boldsymbol{\nabla}\Phi_2)
\end{equation}
if we define the arbitrary amplitude function $g(\xi, \eta)$ in the general solution as
\begin{equation} \label{AmplitudeFunction}
g(\xi, \eta) = (\partial_\xi \Phi_1)(\partial_\eta \Phi_2) - (\partial_\eta \Phi_1)(\partial_\xi \Phi_2)
\end{equation}

Let us note one more relation including the second derivatives with respect to the coordinates and time. Applying $(\partial_t - \partial_z)$ to the first of the equations \eqref{DerivativesRelation},
$(\partial_x - i\partial_y)$ to the second one and then subtracting the results we obtain
\begin{equation*}
(\partial_t^2 - \Delta)\Phi = - (\partial_tv - \partial_zv)(\partial_x\Phi - i\partial_y\Phi) + (\partial_xv - i\partial_yv)(\partial_t\Phi - \partial_z\Phi)
\end{equation*}
From here, using \eqref{Derivatives1} and \eqref{Derivatives2} we simplify the right-hand side and obtain
\begin{equation} \label{SecondDerivatives}
(\partial_t^2 - \Delta)\Phi = \frac{4}{R} (f_\xi \Phi_\eta - f_\eta \Phi_\xi)
\end{equation}
In particular, for the generating function $v(\textbf{r}, t) = f(\xi, \eta)$, the right-hand side is zero, so $v$ satisfies a homogeneous wave equation.
\begin{equation*}
(\partial_t^2 - \Delta)v = 0
\end{equation*}

\section{Bateman's construction}

Comparing \eqref{BatemanField1} and \eqref{BatemanField2} one gets
\begin{equation} \label{bat}
(\partial_t \Phi_2) (\boldsymbol{\nabla} \Phi_1) - (\partial_t \Phi_1) (\boldsymbol{\nabla} \Phi_2) = i(\boldsymbol{\nabla} \Phi_1) \times (\boldsymbol{\nabla} \Phi_2)
\end{equation}
for each pair of arbitrary functions $\Phi_1(\xi, \eta)$ and $\Phi_2(\xi, \eta)$.
The equation \eqref{BatemanField1}, or \eqref{BatemanField2}, together with the equation \eqref{bat} represent the famous Bateman's construction for finding null electromagnetic fields. Bateman$^{[2]}$ has shown that if $\Phi_1$ and $\Phi_2$ are two arbitrary functions that satisfy the condition \eqref{bat}, then the vector \textbf{F} defined by \eqref{BatemanField1}, or \eqref{BatemanField2}, is a null electromagnetic field. Hogan$^{[3]}$ has shown that the Bateman condition \eqref{bat} is also necessary. Our calculations simply confirm the necessity of the condition. Bateman condition is very useful in establishing the properties of the null field  and we use it a couple of times in the following sections.

\section{Natural vector potential}

We have got the general solution to the null electromagnetic field problem, so we do not need a vector potential as a mathematical tool in solving the equations. But if we want to study the behaviour of a Dirac particle in a null electromagnetic field we do need a vector potential to insert into the Dirac equation.

\vspace{1 mm}
\textbf{Definition}
\vspace{1 mm}

According to the standard definition the vector potential is a four-vector 
$A = (A_0, \textbf{A})$, which derivatives give the electric and magnetic field by the expressions
\begin{equation*}
\frac{1}{c} \textbf{E} = -\partial_t \textbf{A} - \boldsymbol{\nabla} A_0 \hspace{10 mm}
\textbf{B} = \boldsymbol{\nabla} \times \textbf{A}
\end{equation*}
This can be written in an equivalent complex form as
\begin{equation} \label{FfromA}
\frac{1}{c} \textbf{F} = -\partial_t \textbf{A} - \boldsymbol{\nabla} A_0
+ i\boldsymbol{\nabla} \times \textbf{A}
\end{equation}
Usually one assumes that A is a real polar vector, it change its sign under spatial inversion.
The Bateman representation of the null electromagnetic \eqref{BatemanField1} and \eqref{BatemanField2} allows to easily find a complex four-vector satisfying \eqref{FfromA}. We can declare the arbitrary functions $\Phi_1(\xi, \eta)$ and $\Phi_2(\xi, \eta)$ to be Lorentz scalars and define the vector potential as
\begin{align} \label{FirstNVP}
A^\mu = \frac{1}{2c} \Phi_2 \partial^\mu \Phi_1 = \frac{1}{2c}(\Phi_2 \partial_t \Phi_1, -\Phi_2\boldsymbol{\nabla}\Phi_1)
\end{align}
Using the Bateman condition \eqref{bat} we obtain from \eqref{FirstNVP}
\begin{align*}
& - \partial_t \textbf{A} - \boldsymbol{\nabla} A_0 + i\boldsymbol{\nabla} \times \textbf{A} 
= \frac{1}{2c} [\ \partial_t(\Phi_2\boldsymbol{\nabla}\Phi_1) - \boldsymbol{\nabla}(\Phi_2\partial_t\Phi_1) - i\boldsymbol{\nabla} \times (\Phi_2\boldsymbol{\nabla}\Phi_1)\ ] \\
& = \frac{1}{2c} [\ (\partial_t\Phi_2)(\boldsymbol{\nabla}\Phi_1) + \Phi_2\boldsymbol{\nabla}\partial_t \Phi_1 - (\boldsymbol{\nabla}\Phi_2)(\partial_t \Phi_1) - \Phi_2\boldsymbol{\nabla}\partial_t \Phi_1 - i(\boldsymbol{\nabla}\Phi_2) \times (\boldsymbol{\nabla}\Phi_1)\ ] \\
& = \frac{1}{2c} [\ (\partial_t \Phi_2)(\boldsymbol{\nabla}\Phi_1) - (\partial_t \Phi_1)(\boldsymbol{\nabla}\Phi_2) + i(\boldsymbol{\nabla}\Phi_1) \times (\boldsymbol{\nabla}\Phi_2)\ ] \\
& = \frac{1}{c} [\ (\partial_t \Phi_2)(\boldsymbol{\nabla}\Phi_1) - (\partial_t \Phi_1)(\boldsymbol{\nabla}\Phi_2)\ ] = \frac{i}{c} (\boldsymbol{\nabla}\Phi_1) \times (\boldsymbol{\nabla}\Phi_2) = \frac{1}{c} \textbf{F}
\end{align*}
The functions $\Phi_1(\xi, \eta)$ and $\Phi_2(\xi, \eta)$ can be considered as complex Euler potentials. The definition and properties of Euler potentials are discussed in detail by Stern$^{[4]}$. The Euler potentials are usualy denoted by $\alpha$ and $\beta$, the vector potential is defined as $\textbf{A} = \alpha \boldsymbol{\nabla} \beta$, and the magnetic field is obtained as $\textbf{B} = \boldsymbol{\nabla} \times \textbf{A} = \boldsymbol{\nabla}\alpha \times \boldsymbol{\nabla}\beta$.
We used \eqref{FirstNVP} as a definition of the vector potential to show that the the arbitrary functions $\Phi_1$ and $\Phi_2$ correspond to the standard Euler potentials. It is clear, however, that the definition of the vector potential through Euler potentials is not unique. If $(\Phi_1, \Phi_2)$ is a pair of functions that generates a given null field, it follows from \eqref{AmplitudeFunction} that any other pair of functions $(\Phi_1^\prime, \Phi_2^\prime)$ will generate the same field, if it satisfy the condition
\begin{equation} \label{PairEquivalenceCondition}
(\partial_\xi \Phi_1^\prime)(\partial_\eta \Phi_2^\prime) - (\partial_\eta \Phi_1^\prime)(\partial_\xi \Phi_2^\prime) =
(\partial_\xi \Phi_1)(\partial_\eta \Phi_2) - (\partial_\eta \Phi_1)(\partial_\xi \Phi_2)
\end{equation}
The functions $\Phi_1$ and $\Phi_2$ are arbitrary and we can always redefine them. In the general case the redefinition has the form
\begin{equation*}
\Phi_1^\prime = \Phi_1^\prime(\Phi_1, \Phi_2) \hspace{10 mm}
\Phi_2^\prime = \Phi_2^\prime(\Phi_1, \Phi_2)
\end{equation*}
It follows from \eqref{PairEquivalenceCondition} that the two pairs of functions will generate the same field if the Jacobian of the transformation is equal to one.
\begin{equation*}
\frac{\partial(\Phi_1^\prime, \Phi_2^\prime)}{\partial(\Phi_1, \Phi_2)} = 1
\end{equation*}
This result is identical to that for standard Euler potentials.

Using this freedom we can define vector potential to be symmetric with respect to Euler potentials.
\begin{equation} \label{deffNVP}
A^{\mu} = \frac{1}{4c}(\Phi_2 \partial^{\mu} \Phi_1 - \Phi_1 \partial^{\mu} \Phi_2)
\hspace{8 mm}
A = \frac{1}{4c}(\Phi_2 \partial_t \Phi_1 - \Phi_1 \partial_t \Phi_2,
-\Phi_2 \boldsymbol{\nabla} \Phi_1 + \Phi_1 \boldsymbol{\nabla} \Phi_2)
\end{equation}
\begin{equation*}
\textbf{F} = \frac{1}{2}
[\ (\partial_t \Phi_2)(\boldsymbol{\nabla} \Phi_1)
- (\partial_t \Phi_1)(\boldsymbol{\nabla} \Phi_2)
+ i (\boldsymbol{\nabla} \Phi_1) \times (\boldsymbol{\nabla} \Phi_2)\ ]
\end{equation*}
The factor $1/4c$ is chosen such that \eqref{FfromA} reproduces \eqref{BatemanField1} and \eqref{BatemanField2}.

I call this vector \emph{natural vector potential} of the null electromagnetic field.

Let $A_\textsc{E} = (A_\textsc{E0}, \textbf{A}_\textsc{E})$ and $A_\textsc{B} = (A_\textsc{B0}, \textbf{A}_\textsc{B})$ be the real and imaginary part of the natural vector potential A.
\begin{equation*}
A_0 = A_\textsc{E0} + i A_\textsc{B0} \hspace{10 mm} \textbf{A} = \textbf{A}_\textsc{E} + i\textbf{A}_\textsc{B}
\end{equation*}
For the electric and magnetic field from \eqref{FfromA} we obtain
\begin{equation} \label{CabibboFerrari}
\frac{1}{c} \textbf{E} = - \partial_t \textbf{A}_\textsc{E} - \boldsymbol{\nabla} A_\textsc{E0} - \boldsymbol{\nabla} \times \textbf{A}_\textsc{B}
\end{equation}
\begin{equation*}
\textbf{B} = - \partial_t \textbf{A}_\textsc{B} - \boldsymbol{\nabla} A_\textsc{B0} + \boldsymbol{\nabla} \times \textbf{A}_\textsc{E}
\end{equation*}
The two real potentials $A_\textsc{E}$ and $A_\textsc{B}$ are usually referred to as Cabibbo and Ferrari$^{[5]}$ potentials.
However, it should be noted that Shanmugadhasan$^{[6]}$ defined them earlier. People who use two real potentials in their studies are usually motivated by an interest in magnetic monopoles or in the duality of the electromagnetic field, or both. Here, we study an electromagnetic field without sources (neither electric nor magnetic) and without assuming anything about duality. The two real potentials appear as a natural part of the general solution.

Let us just note that it is not absolutely necessary that Euler potentials be Lorentz scalars. One can define them as components of a two-component spinor. A left-right spinor pair $(\Psi, \Phi)$ can be used to define a four-vector $A^\mu = \Psi^ \dagger i\partial^\mu \Phi + (\Phi^ \dagger i\partial^\mu \Psi)^*$. Then, reducing the number of the variables by imposing the Lorentz invariant condition $\Psi = -i \sigma_y \Phi^*$, one arrives at the definition \eqref{deffNVP}. The tensor nature of the Euler potentials do not affect the rest of our study and we are not going deeper in this. 

The components of the natural vector potential are not independent. There are  algebraic relations between them and the family generating function. From definition \eqref{deffNVP} and taking into  account \eqref{DerivativesRelation} we obtain
\begin{equation} \label{NVPComponentsRelation}
A_0 - A_z - v(A_x - iA_y) = 0
\end{equation}
\begin{equation*}
A_x + iA_y - v(A_0 + A_z) = 0
\end{equation*}

\vspace{1 mm}
\textbf{Properties}
\vspace{1 mm}

Let us list some of the basic properties of the natural vector potential.
\vspace{2 mm}

1. The natural vector potential satisfies Bateman condition.

From the definition \eqref{deffNVP} we obtain
\begin{equation*}
\partial_t \textbf{A} + \boldsymbol{\nabla} A_0 = \frac{1}{2c}
[\ (\partial_t \Phi_1)(\boldsymbol{\nabla} \Phi_2) - (\partial_t \Phi_2)(\boldsymbol{\nabla} \Phi_1)\ ]
\hspace{8 mm}
\boldsymbol{\nabla} \times \textbf{A} = 
\frac{1}{2c} (\boldsymbol{\nabla} \Phi_1) \times (\boldsymbol{\nabla} \Phi_2)
\end{equation*}
and according to \eqref{bat}
\begin{equation} \label{bat2}
\partial_t \textbf{A} + \boldsymbol{\nabla} A_0
+ i\boldsymbol{\nabla} \times \textbf{A} = 0
\end{equation}
This equation is the Bateman condition in terms of natural vector potential. It is simply \eqref{bat}, written in another form, and we have to keep calling it Bateman condition.
Note that a gauge transformation of the natural vector potential
$A^{\mu} \rightarrow A^{\mu} + \partial^{\mu}\Phi$, where $\Phi$ is an arbitrary function of the coordinates and time, do not change the two expressions
$\partial_t \textbf{A} + \boldsymbol{\nabla} A_0$ and $\boldsymbol{\nabla} \times \textbf{A}$.
Therefore the Bateman condition \eqref{bat2} is true for any potential that is obtained from a natural one by a gauge transformation.
Bateman condition \eqref{bat2} written for the two real potentials is
\begin{equation} \label{bat3}
\partial_t \textbf{A}_\textsc{E} + \boldsymbol{\nabla} A_\textsc{E0} 
- \boldsymbol{\nabla} \times \textbf{A}_\textsc{B} = 0
\end{equation}
\begin{equation*}
\partial_t \textbf{A}_\textsc{B} + \boldsymbol{\nabla} A_\textsc{B0} 
+ \boldsymbol{\nabla} \times \textbf{A}_\textsc{E} = 0
\end{equation*}
Using \eqref{CabibboFerrari} and \eqref{bat3}, we can write two more representations of the electric and magnetic field.
\begin{equation} \label{EBfromAEB}
\frac{1}{2c} \textbf{E} = - \partial_t \textbf{A}_\textsc{E} - \boldsymbol{\nabla} A_\textsc{E0} 
\hspace{8 mm}
\frac{1}{2} \textbf{B} = \boldsymbol{\nabla} \times \textbf{A}_\textsc{E}
\end{equation}
\begin{equation*}
\frac{1}{2} \textbf{B} =- \partial_t \textbf{A}_\textsc{B} - \boldsymbol{\nabla} A_\textsc{B0}
\hspace{8 mm}
\frac{1}{2c} \textbf{E} =  -\boldsymbol{\nabla} \times \textbf{A}_\textsc{B}
\end{equation*}
Clearly, the doubled real part $2A_\textsc{E}$ can be viewed as a conventional real vector potential.
\vspace{2 mm}

2. The natural potential is a substantially complex vector.

Indeed, suppose $A_\textsc{B} = 0$. Then it follows from the Bateman condition \eqref{bat3} that
$\partial_t \textbf{A}_\textsc{E} + \boldsymbol{\nabla} A_\textsc{E0} = 0$ and $\boldsymbol{\nabla} \times \textbf{A}_\textsc{E}$. Thus according to \eqref{EBfromAEB} the field is identically zero.
\vspace{2 mm}

3. The real and imaginary part of the natural potential are not connected by a gauge transformation.

Indeed, suppose that there exists a scalar function $\Phi$, such that 
$A_\textsc{B0} = A_\textsc{E0} + \partial_t \Phi$ and $\textbf{A}_\textsc{B} = \textbf{A}_\textsc{E} - \boldsymbol{\nabla} \Phi$. Then, by replacing these expressions in the Bateman condition \eqref{bat3} we obtain 
$\partial_t \textbf{A}_\textsc{E} + \boldsymbol{\nabla} A_\textsc{E0} - \boldsymbol{\nabla} \times \textbf{A}_\textsc{E} = 0$ 
and 
$\partial_t \textbf{A}_\textsc{E} + \boldsymbol{\nabla} A_\textsc{E0} + \boldsymbol{\nabla} \times \textbf{A}_\textsc{E} = 0$. 
It follows that $\partial_t \textbf{A}_\textsc{E} + \boldsymbol{\nabla} A_\textsc{E0} = 0$ and $\boldsymbol{\nabla} \times \textbf{A}_\textsc{E}$ and, according to \eqref{EBfromAEB}, the field is identically zero.
\vspace{2 mm}

4. The Lorentz gauge condition $\partial_\mu A^{\mu} = 0$
is not true for the natural vector potential in the general case.

From the definition \eqref{deffNVP} we obtain
\begin{equation*}
\partial_t A_0 + \boldsymbol{\nabla}\textbf{.A} = \frac{1}{4c}
[\ \Phi_2(\partial_t^2 - \Delta)\Phi_1 - \Phi_1(\partial_t^2 - \Delta)\Phi_2\ ]
\end{equation*}
and taking into account \eqref{SecondDerivatives}
\begin{equation*}
\partial_t A_0 + \boldsymbol{\nabla}\textbf{.A} = \frac{1}{cR} 
[\ f_\xi (\Phi_2 \partial_\eta \Phi_1 - \Phi_1 \partial_\eta \Phi_2) 
- f_\eta (\Phi_2 \partial_\xi \Phi_1 - \Phi_1 \partial_\xi \Phi_2)\ ]
\end{equation*}

The right-hand side depends on the choice of the family generator $f(\xi, \eta)$ and it is not zero in the general case.
\vspace{2 mm}

5. The natural vector potential satisfies the Dirac$^{[5]}$ gauge condition with a zero mass term on the right-hand side.

Multiplying the first of the two equations \eqref{NVPComponentsRelation} by 
$(A_0 + A_z)$, the second by$(A_x - iA_y)$ and subtracting the resulting equations we get
\begin{equation} \label{DiracGauge1}
A_0^2 - \textbf{A}^2 = 0
\end{equation}
or for the two real potentials
\begin{equation*}
A_\textsc{E0}^2 - \textbf{A}_\textsc{E}^2 = A_\textsc{B0}^2 - \textbf{A}_\textsc{B}^2
\hspace{10 mm}
A_\textsc{E0} A_\textsc{B0} - \textbf{A}_\textsc{E} \textbf{.} \textbf{A}_\textsc{B} = 0
\end{equation*}

6. The natural vector potential is an eigenvector of the complex field tensor.

Consider four-vector
\begin{equation}
H^\mu = F^{\mu\nu}A_\nu = (\textbf{A.F}, A_0\textbf{F} - i\textbf{A} \times \textbf{F})
\end{equation}
It follows from the Maxwell equations \eqref{Maxw1} that, in the general case of a not null field, $H$ satisfies the continuity equation
\begin{equation*}
\partial_\mu H^\mu = - \frac{1}{c} \textbf{F}^2 + \mu_0 c I^\mu A_\mu
\end{equation*}
In the case of a null field $I = 0$ and $\textbf{F}^2 = 0$, the right-hand side is zero, so $H$ defines the density and the flux of a locally conserved quantity.

Using the spinor $\psi$ and taking into account that $\psi_2 = v \psi_1$ the equation \eqref{NVPComponentsRelation} can be written as
\begin{equation*}
( A_0 - \boldsymbol{\sigma} \textbf{.A} ) \psi = 0
\end{equation*}
multiplying this by $\phi^\dagger\boldsymbol{\sigma}$ we obtain
\begin{equation*}
A_0 \phi^\dagger\boldsymbol{\sigma}\psi - \textbf{A} \phi^\dagger \psi
 - i A \times \phi^\dagger\boldsymbol{\sigma} \psi = 0
\end{equation*}
But $\phi^\dagger \boldsymbol{\sigma} \psi$ is the electromagnetic field, $\phi^\dagger \psi$ is zero under null field condition, so
\begin{equation*}
A_0\textbf{F} - i\textbf{A} \times \textbf{F} = 0
\end{equation*}
whence it follows that also
\begin{equation*}
\textbf{A.F} = 0
\end{equation*}
The four-vector $H$ is not just conserved, it is identically zero.
The eigenvalues $s$ of the field tensor $F^{\mu\nu}$, which are roots of the equation $s^2 = \textbf{F}^2$ are zeros for a null electromagnetic field.
For the two real potentials one has
\begin{align*}
& A_\textsc{E0} \textbf{E} + \textbf{A}_\textsc{E} \times c\textbf{B} - \textbf{A}_\textsc{B0} c\textbf{B} + \textbf{A}_\textsc{B} \times \textbf{E} = 0 &
\textbf{A}_\textsc{E}\textbf{.E} - \textbf{A}_\textsc{B}.c\textbf{B} = 0 \\
& A_\textsc{E0} c\textbf{B} - \textbf{A}_\textsc{E} \times \textbf{E} + A_\textsc{B0} \textbf{E} + \textbf{A}_\textsc{B} \times c\textbf{B} = 0 &
\textbf{A}_\textsc{E}.c\textbf{B} + \textbf{A}_\textsc{B}\textbf{.E} = 0
\end{align*}

\vspace{2 mm}

7. Finally, note another algebraic relation between the natural vector potential and the unit Poynting vector $\textbf{n}$
\begin{equation}
A_0 \textbf{n} - \textbf{A} + i\textbf{n} \times \textbf{A} = 0 \hspace{10 mm}
A_0 - \textbf{n.A} = 0
\end{equation}
Or for the two real potentials
\begin{equation}
A_\textsc{E0} \textbf{n} - \textbf{A}_\textsc{E} - \textbf{n} \times \textbf{A}_\textsc{B} = 0
\hspace{10 mm}
A_\textsc{E0} - \textbf{n.A}_\textsc{E} = 0
\end{equation}
\begin{equation*}
A_\textsc{B0} \textbf{n} - \textbf{A}_\textsc{B} + \textbf{n} \times \textbf{A}_\textsc{E} = 0
\hspace{10 mm}
A_\textsc{B0} - \textbf{n.A}_\textsc{B} = 0
\end{equation*}

Note a special form of the natural vector potential for the case of a plane electromagnetic wave. In this case $v = const$ and according to \eqref{Derivatives1} the derivatives of an arbitrary function $\Phi(\xi, \eta)$ are
\begin{equation*}
\partial_t\Phi = -\partial_\xi \Phi - v\partial_\eta \Phi
\hspace{5 mm}
\partial_x \Phi = v\partial_\xi \Phi + \partial_\eta \Phi
\hspace{5 mm}
\partial_y \Phi = -iv\partial_\xi \Phi + i\partial_\eta \Phi
\hspace{5 mm}
\partial_z\Phi = \partial_\xi \Phi - v\partial_\eta \Phi
\end{equation*}
Putting this in the definition \eqref{deffNVP} one has
\begin{equation} \label{PlaneWaveNVP}
A_0 = \Psi_1 + v\Psi_2 \hspace{10 mm}
\textbf{A} = (v\Psi_1 + \Psi_2, -iv\Psi_1 + i\Psi_2, \Psi_1 - v\Psi_2)
\end{equation}
where $\Psi_1$ and $\Psi_2$ are two new arbitrary functions of $\xi$ and $\eta$ defined as
\begin{equation*}
\Psi_1(\xi, \eta) = \frac{1}{4c}(\Phi_1\partial_\xi\Phi_2 - \Phi_2\partial_\xi\Phi_1)
\hspace{5 mm}
\Psi_2(\xi, \eta) = \frac{1}{4c}(\Phi_1\partial_\eta\Phi_2 - \Phi_2\partial_\eta\Phi_1)
\end{equation*}
We will use this later when solving the problem of a Dirac particle in a plane electromagnetic wave.

\section{Lienard-Wiechert field}

The electromagnetic field of an arbitrary moving charged particle in vacuum is calculated from the Lienard-Wiechert potential$^{[8]}$. We would like to point out an interesting property of this field, which can be compared to the corresponding property of the null electromagnetic field.
For a particle moving along a trajectory $\textbf{r}_0(t)$, the retarded Lienard-Wiechert potential is
\begin{equation} \label{LWpot}
A_0 = \frac{q}{4\pi\epsilon_0c} \frac{1}{R - \boldsymbol{\beta}\textbf{.R}}
\hspace{10 mm}
\textbf{A} = A_0 \boldsymbol{\beta}
\end{equation}
where $\textbf{R} = \textbf{r} - \textbf{r}_0(t), \hspace{1 mm} R = |\textbf{R}|, \hspace{1 mm} \boldsymbol{\beta}(t) = \frac{1}{c}\frac{d}{dt}\textbf{r}_0(t)$
and $q$ is the electric charge of the particle. The values of the variables on the right-hand side of \eqref{LWpot} are taken at the retarded time $t - R/c$.

From \eqref{LWpot} one gets
\begin{equation} \label{LWDiracCondition1}
A_0^2 - \textbf{A}^2 = (1 - \beta^2)A_0^2 = 
\left(\frac{q}{4\pi\epsilon_0c}\right)^2 \frac{1 - \beta^2}{(R - \boldsymbol{\beta}\textbf{.R})^2}
\end{equation}

The electric and magnetic fields calculated from \eqref{LWpot} are
\begin{equation} \label{LWField}
\frac{1}{c} \textbf{E}
= A_0 [\ (1 - \beta^2 + \dot{\boldsymbol{\beta}}\textbf{.R})(\textbf{R} - R\boldsymbol{\beta} )
- R(R - \boldsymbol{\beta}\textbf{.R})\dot{\boldsymbol{\beta}}\ ]
/(R - \boldsymbol{\beta}\textbf{.R})^2
\end{equation}
\begin{equation*}
\textbf{B}
= A_0 [\ (1 - \beta^2 + \dot{\boldsymbol{\beta}}\textbf{.R}) \boldsymbol{\beta} \times \textbf{R} + (R - \boldsymbol{\beta}\textbf{.R})\dot{\boldsymbol{\beta}} \times \textbf{R}\ ]
/(R - \boldsymbol{\beta}\textbf{.R})^2
\end{equation*}
where $\dot{\boldsymbol{\beta}}$ is the acceleration of the particle.
The magnetic field is perpendicular to the electric field everywhere
\begin{equation*}
c\textbf{B} = \frac{\textbf{R}}{R} \times \textbf{E}
\end{equation*}
It follows from \eqref{LWField}
\begin{equation*}
\frac{\textbf{R.E}}{R} = cA_0 \frac{1 - \beta^2}{R - \boldsymbol{\beta}\textbf{.R}}
= \frac{q}{4\pi\epsilon_0} \frac{1 - \beta^2}{(R - \boldsymbol{\beta}\textbf{.R})^2}
\end{equation*}
For the complex scalar invariant, we have
\begin{equation*}
\textbf{F}^2 = \textbf{E}^2 - c^2\textbf{B}^2 + 2ic\textbf{E.B} 
= \textbf{E}^2 - \frac{(\textbf{R} \times \textbf{E})^2}{R^2} = \frac{(\textbf{R.E})^2}{R^2}
= \left(\frac{q}{4\pi\epsilon_0}\right)^2
\frac{(1 - \beta^2)^2}{(R - \boldsymbol{\beta}\textbf{.R})^4}
\end{equation*}
Comparing this equation with \eqref{LWDiracCondition1}, we arrive at
\begin{equation*}
(A_0^2 - \textbf{A}^2)^2 = \left(\frac{q}{4\pi\epsilon_0c}\right)^2 \textbf{F}^2
\end{equation*}
or
\begin{equation} \label{ModDiracGauge}
A_0^2 - \textbf{A}^2 = \pm \frac{q}{4\pi\epsilon_0c}\sqrt{\textbf{F}^2}
\end{equation}
The equation \eqref{ModDiracGauge} is a generalization of the corresponding property of the null electromagnetic field \eqref{DiracGauge1}. Obviously \eqref{DiracGauge1} is obtained from \eqref{ModDiracGauge} when the square of the field is zero.
In his 1951 article$^{[5]}$, Dirac proposed to remove the ambiguity in the definition of vector potential by using the condition
\begin{equation*}
A_0^2 - \textbf{A}^2 = k^2
\end{equation*}
where $k$ is a fundamental constant. This condition is called Dirac gauge condition, although the theory itself does not have a significant development. With due respect to the great minds, we keep calling the equation \eqref{ModDiracGauge} Dirac condition, although the right-hand side here is not a constant.
So, the Dirac condition is satisfied for a large set of solutions of the Maxwell equations. This set includes the two model cases "pure radiation" and "isolated charged particle", which are intuitively completely different, and in a classical sense they are incompatible.

\section{Dirac particle in an electromagnetic field}

To describe the behaviour of a Dirac particle in a given electromagnetic field, one has to choose a vector potential that generates this field by the definition \eqref{FfromA}, to put this potential in Dirac equation, and finally solve the resulting equation. We have seen that for a null electromagnetic field, there are two real potentials $A_\textsc{E}$ and $A_\textsc{B}$, which produce one and the same field according to \eqref{EBfromAEB}. These potentials are both non zero, and are not connected to each other by a gauge transformation. They are closely related to each other but not identical.

On the one hand, we have no formal mathematical reason to choose one or the other. On the other hand, we do not have a well-established generally accepted physical interpretation of the vector potential, and we cannot make a choice based on physical arguments. One obvious way out of the problem is just to use the real part $A_\textsc{E}$ and completely ignore the imaginary $A_\textsc{B}$. If we decide to do this we should formulate explicitly some kind of rule, since it would be a hidden postulate in the theory. This approach formally goes around the problem and it is not quite interesting since it gives us nothing new.

There is another option that will be considered here. One can modify the coupling term in the Dirac equation and to make the equation work with both potentials. It is not difficult to do and in my opinion this option should at least be considered.

The Dirac equation in chiral (or Weyl) representation with a real vector potential $A$ is written as
\begin{equation} \label{Dirac1}
(\partial_t + \boldsymbol{\sigma}\textbf{.}\boldsymbol{\nabla}) \phi_{_\textsc{R}} =
-im\phi_{_\textsc{L}} - ie(A_0 - \textbf{A}\textbf{.}\boldsymbol{\sigma}) \phi_{_\textsc{R}}
\end{equation}
\begin{equation*}
(\partial_t - \boldsymbol{\sigma}\textbf{.}\boldsymbol{\nabla}) \phi_{_\textsc{L}} =
-im\phi_{_\textsc{R}} - ie(A_0 + \textbf{A} \textbf{.}\boldsymbol{\sigma}) \phi_{_\textsc{L}}
\end{equation*}
where
$m = m_0c/\hbar$, $e = q/\hbar$
for a particle with rest mass $m_0$ and electrical charge $q$. $\phi_{_\textsc{R}}$ and $\phi_{_\textsc{L}}$ are one right-handed and one left-handed two-component Weyl spinors respectively. These spinors are connected with the two-component Dirac spinors $\psi_{_\textsc{A}}$ and $\psi_{_\textsc{B}}$ as follows.
\begin{equation*}
\phi_{_\textsc{R}} = (\psi_{_\textsc{A}} + \psi_{_\textsc{B}})/\sqrt{2} \hspace{20 mm}
\phi_{_\textsc{L}} = (\psi_{_\textsc{A}} - \psi_{_\textsc{B}})/\sqrt{2}
\end{equation*}
The chiral representation suits our purpose, since the equation \eqref{Dirac1} is manifestly covariant under spatial inversion.
Let us define one right-handed $A_\textsc{R} = (A_\textsc{R0}, \textbf{A}_\textsc{R})$ and one left-handed $A_\textsc{L} = (A_\textsc{L0}, \textbf{A}_\textsc{L})$ chiral potentials as follows.
\begin{equation}
A_\textsc{R0} = A_\textsc{E0} + A_\textsc{B0} \hspace{5 mm} 
\textbf{A}_\textsc{R} = \textbf{A}_\textsc{E} + \textbf{A}_\textsc{B}
\hspace{10 mm}
A_\textsc{L0} = A_\textsc{E0} - A_\textsc{B0} \hspace{5 mm} 
\textbf{A}_\textsc{L} = \textbf{A}_\textsc{E} - \textbf{A}_\textsc{B}
\end{equation}
Since $A_\textsc{E}$ is a polar and $A_\textsc{B}$ is an axial vector, under spatial inversion, chiral potentials are transformed as
$A_\textsc{R0} \rightarrow A_\textsc{L0}$, $\textbf{A}_\textsc{R} \rightarrow -\textbf{A}_\textsc{L}$, $A_\textsc{L0} \rightarrow A_\textsc{R0}$, $\textbf{A}_\textsc{L} \rightarrow -\textbf{A}_\textsc{R}$. 
We can put the right-handed potential $A_\textsc{R}$ in the right-handed equation, and the left-handed potential $A_\textsc{L}$ in the left-handed equation \eqref{Dirac1}. We arrive at
\begin{equation} \label{Dirac2}
(\partial_t + \boldsymbol{\sigma}\textbf{.}\boldsymbol{\nabla}) \phi_{_\textsc{R}} =
-im\phi_{_\textsc{L}} - ie(A_\textsc{R0} - \textbf{A}_\textsc{R}\textbf{.}\boldsymbol{\sigma}) \phi_{_\textsc{R}}
\end{equation}
\begin{equation*}
(\partial_t - \boldsymbol{\sigma}\textbf{.}\boldsymbol{\nabla}) \phi_{_\textsc{L}} =
-im\phi_{_\textsc{R}} - ie(A_\textsc{L0} + \textbf{A}_\textsc{L}\textbf{.}\boldsymbol{\sigma}) \phi_{_\textsc{L}}
\end{equation*}
This is perhaps the simplest possible modification of the coupling term, such that both real potentials are in use.
The equation \eqref{Dirac2} has the same basic symmetries as the original equation.
It remains unchanged under spatial inversion
\begin{equation*}
\boldsymbol{\nabla} \rightarrow -\boldsymbol{\nabla}, \hspace{2 mm}
\phi_{_\textsc{R}} \rightarrow \phi_{_\textsc{L}}, \hspace{2 mm} 
\phi_{_\textsc{L}} \rightarrow \phi_{_\textsc{R}}
\end{equation*}
\begin{equation*}
A_\textsc{R0} \rightarrow A_\textsc{L0}, \hspace{2 mm}
\textbf{A}_\textsc{R} \rightarrow -\textbf{A}_\textsc{L}, \hspace{2 mm}
A_\textsc{L0} \rightarrow A_\textsc{R0}, \hspace{2 mm}
\textbf{A}_\textsc{L} \rightarrow -\textbf{A}_\textsc{R} 
\end{equation*}
under charge conjugation
\begin{equation*}
\phi_{_\textsc{R}} \rightarrow i\sigma_y \phi_{_\textsc{L}}^*, \hspace{2 mm} \phi_{_\textsc{L}} \rightarrow -i\sigma_y \phi_{_\textsc{R}}^*
\end{equation*}
\begin{equation*}
A_\textsc{R0} \rightarrow -A_\textsc{L0}, \hspace{2 mm} 
\textbf{A}_\textsc{R} \rightarrow -\textbf{A}_\textsc{L} \hspace{5 mm}
A_\textsc{L0} \rightarrow -A_\textsc{R0}, \hspace{2 mm} 
\textbf{A}_\textsc{L} \rightarrow -\textbf{A}_\textsc{R}
\end{equation*}
and is still gauge invariant
\begin{equation*}
\phi_{_\textsc{R}} \rightarrow \phi_{_\textsc{R}} e^{ie\Phi}, \hspace{2 mm} \phi_{_\textsc{L}} \rightarrow \phi_{_\textsc{L}} e^{ie\Phi}
\end{equation*}
\begin{equation*}
A_\textsc{R0} \rightarrow A_\textsc{R0} + \partial_t \Phi, \hspace{2 mm} \textbf{A}_\textsc{R} \rightarrow \textbf{A}_\textsc{R} - \boldsymbol{\nabla} \Phi \hspace{5 mm}
A_\textsc{L0} \rightarrow A_\textsc{L0} + \partial_t \Phi, \hspace{2 mm} \textbf{A}_\textsc{L} \rightarrow \textbf{A}_\textsc{L} - \boldsymbol{\nabla} \Phi
\end{equation*}

Now we can return to Dirac spinors and potentials $A_\textsc{E}$ and $A_\textsc{B}$.
\begin{equation} \label{Dirac3}
\partial_t \psi_{_\textsc{A}} + \boldsymbol{\sigma}\textbf{.}\boldsymbol{\nabla} \psi_{_\textsc{B}} =
-im\psi_{_\textsc{A}} - ie(A_\textsc{E0}\psi_{_\textsc{A}} 
- \textbf{A}_\textsc{E}\textbf{.}\boldsymbol{\sigma}\psi_{_\textsc{B}}) 
- ie(A_\textsc{B0}\psi_{_\textsc{B}} 
- \textbf{A}_\textsc{B}\textbf{.}\boldsymbol{\sigma}\psi_{_\textsc{A}}) 
\end{equation}
\begin{equation*}
\partial_t \psi_{_\textsc{B}} + \boldsymbol{\sigma}\textbf{.}\boldsymbol{\nabla} \psi_{_\textsc{A}} =
im\psi_{_\textsc{B}} - ie(A_\textsc{E0}\psi_{_\textsc{B}} 
- \textbf{A}_\textsc{E}\textbf{.}\boldsymbol{\sigma}\psi_{_\textsc{A}}) 
- ie(A_\textsc{B0}\psi_{_\textsc{A}} 
- \textbf{A}_\textsc{B}\textbf{.}\boldsymbol{\sigma}\psi_{_\textsc{B}})
\end{equation*}
This equation differs from the original Dirac equation only by the term containing $A_B$. This term is zero when the potential is real. So all the results we have or that we could have with a real potential remain unchanged.

Using a coupling term, that contains a pseudovector potential in the Dirac equation, is not something completely new. Lochak$^{[9]}$ proposed a modification of the Dirac equation that contains a similar term. Lochak's aim was to write a wave equation that describes a massless magnetic monopole. Accordingly, the coupling term in his equation is proportional to the magnetic charge of the monopole. While here, we consider a conventional massive electrically charged particle that interacts with an electromagnetic field and "feels" both real potentials.

\section{Volkov type solution for a Dirac particle \\in a plane electromagnetic wave}
The solution of the Dirac equation for a charged particle in a plane electromagnetic wave was found by Volkov$^{[10]}$. Here, we will solve the same problem, but with a modified coupling term in the Dirac equation, taking into account both real potentials. We want to directly compare our result to that of Volkov, so we try to be as close as possible to the assumptions that he made to simplify the problem. This will give us an idea of what to expect from the modification of the coupling term in the Dirac equation.

When solving the conventional problem, it is assumed that the real vector potential $A(\zeta)$ depends on the coordinates and time only through the phase
\begin{equation}
\zeta = \textbf{k}_{_\textsc{F}}.\textbf{r} - \omega_{_\textsc{F}} t = \textbf{k}_{_\textsc{F}}\textbf{.r} - k_{_\textsc{F0}} ct = k_{_\textsc{F0}} (\textbf{n.r} - ct)
\end{equation}
where $k_{_\textsc{F}} = (k_{_\textsc{F0}}, \textbf{k}_{_\textsc{F}}) = (\omega_{_\textsc{F}}/c, \textbf{k}_{_\textsc{F}})$ is the wave vector satisfying the condition $k_{_\textsc{F0}}^2 - \textbf{k}_{_\textsc{F}}^2 = 0$.
Under these assumptions one has
\begin{equation*}
\partial_t A_0 + \boldsymbol{\nabla} \textbf{.A} = - \partial_\zeta(k_{_\textsc{F0}} A_0 - \textbf{k}_{_\textsc{F}}\textbf{.A}) \hspace{10 mm}
(\partial_t^2 - \Delta)A = 0
\end{equation*}
\begin{equation*}
\textbf{F}/c = \partial_\zeta (k_{_\textsc{F0}} \textbf{A} - \textbf{k}_{_\textsc{F}} A_0 + i\textbf{k}_{_\textsc{F}} \times \textbf{A}) \hspace{11 mm}
\textbf{F}^2/c^2 = 
[\ \partial_\zeta(k_{_\textsc{F0}} A_0 - \textbf{k}_{_\textsc{F}}\textbf{.A})\ ]^2
\end{equation*}
It follows that if
\begin{equation} \label{VolkovCondition}
k_{_\textsc{F0}} A_0 - \textbf{k}_{_\textsc{F}}\textbf{.A} = 0
\end{equation}
then the Lorentz gauge condition is valid, the Maxwell equations are satisfied and the square of the field is zero. So this vector potential generates a null electromagnetic field.

\vspace{2 mm}

Now we have to write in a general form a natural vector potential matching the above conditions. This is not quite straightforward since, as we have seen, the null electromagnetic field is constructed by arbitrary functions of the variables
$\xi$ and $\eta$.
These variables can be expressed by the unit Poynting vector $\textbf{n}$ as follows.
\begin{equation*}
\xi = [\ \textbf{n.r} - ct + (\textbf{r} - \textbf{n}ct + i\textbf{n} \times \textbf{r})_z\ ] / (1 + n_z)
\end{equation*}
\begin{equation*}
\eta = 
[\ (\textbf{r} - \textbf{n}ct + i\textbf{n} \times \textbf{r})_x + i(\textbf{r} - \textbf{n}ct + i\textbf{n} \times \textbf{r})_y\ ] / (1 + n_z)
\end{equation*}
It is seen that, in the general case, the variables $\xi$ and $\eta$ mix the coordinates and time in such a way that the value $\textbf{n.r} - ct$ corresponding to phase $\zeta$ cannot be considered as a separate variable.
However, this is possible in the special case of a plane wave propagating in a positive direction along the $z$ axis. In this case $v = 0$ and $\xi = z - ct$ and $\eta = x + iy$. The variable $\xi$ can now be considered as an analog of phase $\zeta$.

For the natural vector potential from \eqref{PlaneWaveNVP} with v = 0 we have
\begin{equation*}
A_0 = \Psi_1, \hspace{5 mm}
\textbf{A} = (\Psi_2, i\Psi_2, \Psi_1)
\end{equation*}
where in the general case $\Phi_1$ and $\Phi_2$ are arbitrary complex functions of $\xi$ and $\eta$. Since we need a potential that depends only on the phase, we obviously need to restrict them to depend only on the variable $\xi$. Under this assumption we can always eliminate the function $\Phi_1$ by a gauge transformation.
\begin{align*}
& A_0 \rightarrow A_0 + \partial_t \Phi(\xi) = \Psi_1 - \partial_\xi \Phi \hspace{8 mm}
A_x \rightarrow A_x - \partial_x \Phi(\xi) = \Psi_2 \\
& A_z \rightarrow A_z - \partial_z \Phi(\xi) = \Psi_1 - \partial_\xi \Phi\hspace{8 mm}
A_y \rightarrow A_y - \partial_y \Phi(\xi) = i\Psi_2 
\end{align*}
We just have to choose a gauge function $\Phi(\xi)$ so that $\partial_\xi \Phi = \Psi_1$. 
Since $\xi$ is a real variable we can write the complex function $\Psi_2$ as
$\Psi_2 = \Phi_1 + i\Phi_2$ where $\Phi_1$ and $\Phi_2$ are arbitrary real functions.
The natural vector potential now is
\begin{equation*}
A_0 = 0 \hspace{10 mm}
\textbf{A} = (\Phi_1 + i\Phi_2, i\Phi_1 - \Phi_2, 0)
\end{equation*}
Or for the two real potentials
\begin{align*}
\textbf{A}_\textsc{E} & = (\Phi_1, -\Phi_2, 0) = \Phi_1 (1, 0, 0) - \Phi_2 (0, 1, 0) \\
\textbf{A}_\textsc{B} & = (\Phi_2, \Phi_1, 0) = \Phi_2 (1, 0, 0) + \Phi_1 (0, 1, 0)
\end{align*}
Finally, we can get rid of the assumption that the wave propagates in a positive direction along the $z$ axis by using a spatial rotation of the coordinate system. The natural vector potential can be written as
\begin{equation} \label{SimplePots1}
A_\textsc{E0} = 0 \hspace{10 mm}
\textbf{A}_\textsc{E} = \Phi_1 \textbf{n}_{_\textsc{E}} - \Phi_2 \textbf{n}_{_\textsc{B}}
\end{equation}
\begin{equation*}
A_\textsc{B0} = 0 \hspace{10 mm}
\textbf{A}_\textsc{B} = \Phi_2 \textbf{n}_{_\textsc{E}} + \Phi_1 \textbf{n}_{_\textsc{B}}
\end{equation*}
where $\textbf{n}_{_\textsc{E}}$ and $\textbf{n}_{_\textsc{B}}$ are two arbitrary orthogonal real unit vectors. These vectors and unit Poynting vector form orthonormal right-handed triad
\begin{equation*}
\textbf{n}_{_\textsc{E}}^2 = \textbf{n}_{_\textsc{B}}^2 = 1  \hspace{10 mm}
\textbf{n}_{_\textsc{E}} \textbf{.n}_{_\textsc{B}} = 0  \hspace{10 mm}
\textbf{n}_{_\textsc{E}} \times \textbf{n}_{_\textsc{B}} = \textbf{n}
\end{equation*}
The corresponding field is
\begin{equation*}
\textbf{E} = 2\omega_{_\textsc{F}}( \partial_\zeta \Phi_1 \textbf{n}_{_\textsc{E}} 
- \partial_\zeta \Phi_2 \textbf{n}_{_\textsc{B}}) \hspace{10 mm}
c\textbf{B} = 2\omega_{_\textsc{F}}( \partial_\zeta \Phi_2 \textbf{n}_{_\textsc{E}} 
+ \partial_\zeta \Phi_1 \textbf{n}_{_\textsc{B}})
\end{equation*}

\vspace{2 mm}

Following the original idea of the Volkov's solution, we assume that the solution of the Dirac equation \eqref{Dirac2} can be represented as a modified plane wave.
\begin{equation} \label{modwave}
\phi_{_\textsc{R}} = \chi_{_\textsc{R}}(\zeta)e^{i(\textbf{k}_{_\textsc{P}}\textbf{.r}\ -\ \omega_{_\textsc{P}} t)} \hspace{10 mm}
\phi_{_\textsc{L}} = \chi_{_\textsc{L}}(\zeta)e^{i(\textbf{k}_{_\textsc{P}}\textbf{.r}\ -\ \omega_{_\textsc{P}} t)}
\end{equation}
where the new unknown spinors $\chi_{_\textsc{R}}$ and $\chi_{_\textsc{L}}$ depend only on the phase $\zeta$. The vector $k_{_\textsc{P}} = (k_{_\textsc{P0}}, \textbf{k}_{_\textsc{P}}) = (\omega_{_\textsc{P}}/c, \textbf{k}_{_\textsc{P}})$ is a wave vector of a free particle and satisfies the condition
\begin{equation*}
k_{_\textsc{P0}}^2 - \textbf{k}_{_\textsc{P}}^2 = m^2
\end{equation*}
By inserting \eqref{modwave} into Dirac equation \eqref{Dirac2} we obtain
\begin{equation} \label{dirac3}
(k_{_\textsc{F0}} - \textbf{k}_{_\textsc{F}}.\boldsymbol{\sigma}) \partial_\zeta \chi_{_\textsc{R}} +
i(k_{_\textsc{P0}} - \textbf{k}_{_\textsc{P}}\textbf{.}\boldsymbol{\sigma})\chi_{_\textsc{R}} = 
im\chi_{_\textsc{L}} 
+ ie(A_\textsc{R0} - \textbf{A}_\textsc{R}\textbf{.}\boldsymbol{\sigma}) \chi_{_\textsc{R}}
\end{equation}
\begin{equation*}
(k_{_\textsc{F0}} + \textbf{k}_{_\textsc{F}}.\boldsymbol{\sigma}) \partial_\zeta \chi_{_\textsc{L}} +
i(k_{_\textsc{P0}} + \textbf{k}_{_\textsc{P}}\textbf{.}\boldsymbol{\sigma})\chi_{_\textsc{L}} 
= im\chi_{_\textsc{R}} 
+ ie(A_\textsc{L0} + \textbf{A}_\textsc{L}\textbf{.}\boldsymbol{\sigma}) \chi_{_\textsc{L}}
\end{equation*}
Multiplying the first equation by $(k_{_\textsc{F0}} + \textbf{k}_{_\textsc{F}}.\boldsymbol{\sigma})$, the second one by $(k_{_\textsc{F0}} - \textbf{k}_{_\textsc{F}}.\boldsymbol{\sigma})$, and taking into account that
\begin{equation*}
(k_{_\textsc{F0}} \pm \textbf{k}_{_\textsc{F}} \textbf{.}\boldsymbol{\sigma})(k_{_\textsc{F0}} \mp \textbf{k}_{_\textsc{F}} \textbf{.}\boldsymbol{\sigma})
= k_{_\textsc{F0}}^2 - \textbf{k}_{_\textsc{F}}^2 = 0
\end{equation*}
\begin{equation*}
(k_{_\textsc{F0}} \pm \textbf{k}_{_\textsc{F}} \textbf{.}\boldsymbol{\sigma})(A_\textsc{RL0} \mp \textbf{A}_\textsc{RL}\textbf{.}\boldsymbol{\sigma}) +
(A_\textsc{RL0} \pm \textbf{A}_\textsc{RL}\textbf{.}\boldsymbol{\sigma})(k_{_\textsc{F0}} \mp \textbf{k}_{_\textsc{F}}\textbf{.}\boldsymbol{\sigma}) = 
2(k_{_\textsc{F0}} A_\textsc{RL0} - \textbf{k}_{_\textsc{F}} \textbf{.A}_\textsc{RL}) = 0
\end{equation*}
we obtain
\begin{equation} \label{relat}
2a \chi_{_\textsc{R}} = m(k_{_\textsc{F0}} + \textbf{k}_{_\textsc{F}} \textbf{.}\boldsymbol{\sigma})\chi_{_\textsc{L}}
+ [\ k_{_\textsc{P0}} + \textbf{k}_{_\textsc{P}} \textbf{.}\boldsymbol{\sigma}
- e(A_\textsc{R0} + \textbf{A}_\textsc{R}\textbf{.}\boldsymbol{\sigma})\ ](k_{_\textsc{F0}} - \textbf{k}_{_\textsc{F}} \textbf{.}\boldsymbol{\sigma})\chi_{_\textsc{R}}
\end{equation}
\begin{equation*} 
2a \chi_{_\textsc{L}} = 
m(k_{_\textsc{F0}} - \textbf{k}_{_\textsc{F}} \textbf{.}\boldsymbol{\sigma})\chi_{_\textsc{R}}
+ [\ k_{_\textsc{P0}} - \textbf{k}_{_\textsc{P}} \textbf{.}\boldsymbol{\sigma}
- e(A_\textsc{L0} - \textbf{A}_\textsc{L}\textbf{.}\boldsymbol{\sigma})\ ](k_{_\textsc{F0}} + \textbf{k}_{_\textsc{F}} \textbf{.}\boldsymbol{\sigma})\chi_{_\textsc{L}}
\end{equation*}
where we have put for brevity
\begin{equation*}
a = k_{_\textsc{F0}} k_{_\textsc{P0}} - \textbf{k}_{_\textsc{F}}\textbf{.k}_{_\textsc{P}}
\end{equation*}
The relation \eqref{relat} between spinors $\chi_{_\textsc{R}}$ and $\chi_{_\textsc{L}}$ actually decreases the number of the unknowns. It is convenient to introduce one single spinor $U$ defined as
\begin{equation*}
2k_{_\textsc{F0}} U =
(k_{_\textsc{F0}} - \textbf{k}_{_\textsc{F}}.\boldsymbol{\sigma})\chi_{_\textsc{R}} +
(k_{_\textsc{F0}} + \textbf{k}_{_\textsc{F}}.\boldsymbol{\sigma})\chi_{_\textsc{L}} 
\end{equation*}
It follows from this definition that
\begin{equation*}
(k_{_\textsc{F0}} - \textbf{k}_{_\textsc{F}}.\boldsymbol{\sigma})\chi_{_\textsc{R}} = 
(k_{_\textsc{F0}} - \textbf{k}_{_\textsc{F}}.\boldsymbol{\sigma})U \hspace{10 mm}
(k_{_\textsc{F0}} + \textbf{k}_{_\textsc{F}}.\boldsymbol{\sigma})\chi_{_\textsc{L}} = 
(k_{_\textsc{F0}} + \textbf{k}_{_\textsc{F}}.\boldsymbol{\sigma})U
\end{equation*}
Putting these expressions in \eqref{relat}, we obtain
\begin{equation} \label{XfromU}
2a \chi_{_\textsc{R}} = m(k_{_\textsc{F0}} + \textbf{k}_{_\textsc{F}}\textbf{.}\boldsymbol{\sigma})U
+ [\ k_{_\textsc{P0}} + \textbf{k}_{_\textsc{P}} \textbf{.}\boldsymbol{\sigma}
- e(A_\textsc{R0} + \textbf{A}_\textsc{R} \textbf{.}\boldsymbol{\sigma})\ ](k_{_\textsc{F0}} - \textbf{k}_{_\textsc{F}} \textbf{.}\boldsymbol{\sigma})U
\end{equation}
\begin{equation*} 
2a \chi_{_\textsc{L}} = m(k_{_\textsc{F0}} - \textbf{k}_{_\textsc{F}} \textbf{.}\boldsymbol{\sigma})U
+ [\ k_{_\textsc{P0}} - \textbf{k}_{_\textsc{P}} \textbf{.}\boldsymbol{\sigma}
- e(A_\textsc{L0} - \textbf{A}_\textsc{L} \textbf{.}\boldsymbol{\sigma})\ ](k_{_\textsc{F0}} + \textbf{k}_{_\textsc{F}}\textbf{.}\boldsymbol{\sigma})U
\end{equation*}

We will write a differential equation for the unknown spinor $U$. Then if we are able to solve this equation the expressions \eqref{XfromU} will give us the solution we are looking for.

We add the second of the equations \eqref{dirac3} to the first one and obtain
\begin{equation*}
2ik_{_\textsc{F0}} \partial_\zeta U = (k_{_\textsc{P0}} - m)(\chi_{_\textsc{R}} + \chi_{_\textsc{L}})
- \textbf{k}_{_\textsc{P}} \textbf{.}\boldsymbol{\sigma}(\chi_{_\textsc{R}} - \chi_{_\textsc{L}})
- e(A_\textsc{R0} - \textbf{A}_\textsc{R} \textbf{.}\boldsymbol{\sigma}) \chi_{_\textsc{R}}
- e(A_\textsc{L0} + \textbf{A}_\textsc{L} \textbf{.}\boldsymbol{\sigma}) \chi_{_\textsc{L}}
\end{equation*}
Here we replace $\chi_{_\textsc{R}}$ and $\chi_{_\textsc{L}}$ with the expressions \eqref{XfromU}. After a straightforward calculation and switching to the potentials $A_\textsc{E}$ and $A_\textsc{B}$, the right-hand side simplifies and we obtain
\begin{equation} \label{DiffEqForU}
ia\partial_\zeta U = 
-e(k_{_\textsc{P0}} A_\textsc{E0} - \textbf{k}_{_\textsc{P}} \textbf{.A}_\textsc{E})U 
+ \frac{e^2}{2}(A_\textsc{E0}^2 - \textbf{A}_\textsc{E}^2 + A_\textsc{B0}^2 - \textbf{A}_\textsc{B}^2)U 
\end{equation}
\begin{equation*}
+ \hspace{1 mm} m e(A_\textsc{B0} \textbf{k}_{_\textsc{F}} - k_{_\textsc{F0}} \textbf{A}_\textsc{B})\textbf{.}\boldsymbol{\sigma}U 
+ e(k_{_\textsc{P0}} A_\textsc{B0} - \textbf{k}_{_\textsc{P}}\textbf{.A}_\textsc{B})U
- e^2(A_\textsc{E0} A_\textsc{B0} - \textbf{A}_\textsc{E}\textbf{.A}_\textsc{B})\textbf{n.}\boldsymbol{\sigma}U
\end{equation*}

The tensors that could be of a physical interest are expressed by the spinor $U$. In particular, the Dirac probability current is
\vspace{1 mm}
\begin{equation} \label{ProbabilityCurrent}
\frac{1}{2}(k_{_\textsc{P0}} - \textbf{n.k}_{_\textsc{P}})^2 J_\textsc{D0} = 
(k_{_\textsc{P0}} - \textbf{n.k}_{_\textsc{P}})(k_{p0} - e A_\textsc{E0})U^\dagger U 
+ (k_{_\textsc{P0}} - \textbf{n.k}_{_\textsc{P}}) \textbf{n} \textbf{.}
U^\dagger \boldsymbol{\sigma} U e A_\textsc{B0}
\end{equation}
\begin{equation*}
\hspace{4 mm} + [\ e(k_{_\textsc{P0}} A_\textsc{E0} - \textbf{k}_{_\textsc{P}} \textbf{.A}_\textsc{E}) 
- \frac{e^2}{2}(A_\textsc{E0}^2 - \textbf{A}_\textsc{E}^2 + A_\textsc{B0}^2 - \textbf{A}_\textsc{B}^2)\ ]U^\dagger U
\end{equation*}
\begin{equation*}
\hspace{4 mm} + m e ( A_\textsc{B0} \textbf{n} - \textbf{A}_\textsc{B} )\textbf{.} U^\dagger \boldsymbol{\sigma} U
- [\ e(k_{_\textsc{P0}} A_\textsc{B0} - \textbf{k}_{_\textsc{P}}\textbf{.A}_\textsc{B}) 
- e^2(A_\textsc{E0} A_\textsc{B0} - \textbf{A}_\textsc{E} \textbf{.A}_\textsc{B})\ ]\textbf{n} \textbf{.}U^\dagger \boldsymbol{\sigma} U
\end{equation*}
\begin{equation*}
\frac{1}{2}(k_{_\textsc{P0}} - \textbf{n.k}_{_\textsc{P}})^2 \textbf{J}_\textsc{D} = 
(k_{_\textsc{P0}} - \textbf{n.k}_{_\textsc{P}})(\textbf{k}_{_\textsc{P}} - e \textbf{A}_\textsc{E}) U^\dagger U 
+ (k_{_\textsc{P0}} - \textbf{n.k}_{_\textsc{P}}) \textbf{n} \textbf{.} U^\dagger \boldsymbol{\sigma} U e \textbf{A}_\textsc{B}
\end{equation*}
\begin{equation*}
\hspace{4 mm} + [\ e(k_{_\textsc{P0}} A_\textsc{E0} - \textbf{k}_{_\textsc{P}}\textbf{.A}_\textsc{E}) 
- \frac{e^2}{2} (A_\textsc{E0}^2 - \textbf{A}_\textsc{E}^2 + A_\textsc{B0}^2 - \textbf{A}_\textsc{B}^2)\ ]U^\dagger U\textbf{n}
\end{equation*}
\begin{equation*}
\hspace{4 mm} + m e ( A_\textsc{B0} \textbf{n} - \textbf{A}_\textsc{B} )\textbf{.} U^\dagger \boldsymbol{\sigma} U \textbf{n}
- [\ e(k_{_\textsc{P0}} A_\textsc{B0} - \textbf{k}_{_\textsc{P}}\textbf{.A}_\textsc{B}) 
- e^2(A_\textsc{E0} A_\textsc{B0} - \textbf{A}_\textsc{E} \textbf{.A}_\textsc{B})\ ]\textbf{n} \textbf{.}U^\dagger \boldsymbol{\sigma} U \textbf{n}
\end{equation*}
The differential equation is simplified further by the substitution $U = V e^{iS}$ with
\begin{equation*}
S = \frac{e}{a} \int d \zeta\ [\ (k_{_\textsc{P0}} A_\textsc{E0} - \textbf{k}_{_\textsc{P}} \textbf{.A}_\textsc{E})
- \frac{e}{2} (A_\textsc{E0}^2 - \textbf{A}_\textsc{E}^2 + A_\textsc{B0}^2 - \textbf{A}_\textsc{B}^2)\ ]
\end{equation*}
The equation for $V$ is
\begin{equation} \label{DiffEqForV}
ia\partial_\zeta V =
m e(A_\textsc{B0} \textbf{k}_{_\textsc{F}} - k_{_\textsc{F0}} 
\textbf{A}_\textsc{B})\textbf{.}\boldsymbol{\sigma}V 
+ e(k_{_\textsc{P0}} A_\textsc{B0} - \textbf{k}_{_\textsc{P}} \textbf{.A}_\textsc{B})V
- e^2(A_\textsc{E0} A_\textsc{B0} - \textbf{A}_\textsc{E} \textbf{.A}_\textsc{B})\textbf{n.}\boldsymbol{\sigma}V
\end{equation}
And using the natural vector potential \eqref{SimplePots1} we obtain
\begin{equation*}
ia\partial_\zeta V = 
e\Phi_1 ( m \textbf{n}_{_\textsc{B}} - \textbf{k}_p \textbf{.n}_{_\textsc{B}} \textbf{n} ) \textbf{.} \boldsymbol{\sigma} V
+ e\Phi_2 ( m \textbf{n}_{_\textsc{E}} - \textbf{k}_p \textbf{.n}_{_\textsc{E}} \textbf{n} ) \textbf{.} \boldsymbol{\sigma} V
\end{equation*}

We want to find at least one solution in a closed easy to read form, so we restrict our study to the simplest case of a linearly polarized electromagnetic wave
\begin{equation} \label{SimplePots2}
\Phi_1 = \Phi \hspace{6 mm} \Phi_2 = 0 \hspace{6 mm}
\textbf{A}_\textsc{E} = \Phi \textbf{n}_{_\textsc{E}} \hspace{6 mm}
\textbf{A}_\textsc{B} = \Phi \textbf{n}_{_\textsc{B}} 
\end{equation}
The equation for $V$ simplifies to
\begin{equation*}
ia\partial_\zeta V = 
e\Phi ( m \textbf{n}_{_\textsc{B}} - \textbf{k}_{_\textsc{P}} \textbf{.n}_{_\textsc{B}} \textbf{n} ) \textbf{.} \boldsymbol{\sigma} V
\end{equation*}
A change of the variable
\begin{equation*}
\eta(\zeta) = \frac{e}{a} \int d \zeta\ \Phi
\end{equation*}
gives an equation with constant coefficients
\begin{equation*}
i \partial_\eta V = MV
\end{equation*}
where M is the constant matrix
\begin{equation*}
M = m \textbf{n}_{_\textsc{B}} \textbf{.} \boldsymbol{\sigma} 
- \textbf{k}_{_\textsc{P}} \textbf{.n}_{_\textsc{B}} \textbf{n} \textbf{.} \boldsymbol{\sigma}
\end{equation*}
The square of this matrix is proportional to the unit matrix. 
Using this we can write an easily solvable second order equation for each of the components of the unknown spinor $V$
\begin{equation*}
\partial_\eta^2 V + \beta^2V = 0
\end{equation*}
where
\begin{equation*}
\beta^2 = M^2 = m^2 + (\textbf{k}_{_\textsc{P}} \textbf{.n}_{_\textsc{B}})^2
\end{equation*}
Finally the spinor V can be written as a linear combination of the solutions corresponding to positive and negative $\beta$ as
\begin{equation*}
V = (\beta - M)W^{(+)} e^{i\beta\eta} + (\beta + M)W^{(-)} e^{-i\beta\eta}
\end{equation*}
where $W^{(+)}$ and $W^{(-)}$ are two arbitrary constant spinors.
The spinor $U$ is
\begin{equation} \label{result1}
U = [\ (\beta - M)W^{(+)} e^{i\beta\eta} + (\beta + M)W^{(-)} e^{-i\beta\eta}\ ]e^{iS}
\end{equation}
Let $U_0$ be the spinor $U$ in the absence of an electromagnetic field
\begin{equation*}
U_0 = U|_{\Phi = 0}
\end{equation*}
When $\Phi = 0$ the expression \eqref{result1} is
\begin{equation*}
U_0 = (\beta - M)W^{(+)} + (\beta + M)W^{(-)}
\end{equation*}
and given that $M^2 = \beta^2$
\begin{equation*}
2\beta (\beta - M) W^{(+)} = (\beta - M) U_0 \hspace{10 mm}
2\beta (\beta + M) W^{(-)} = (\beta + M) U_0
\end{equation*}
We can replace these expressions in \eqref{result1}. Thus, the solution is expressed by one arbitrary constant spinor $U_0$ instead of two $W^{(+)}$ and $W^{(-)}$.
\begin{equation} \label{EquationForU}
\beta U = [\ \beta \cos(\beta\eta) - iM \sin(\beta\eta)\ ] U_0 e^{iS}
\end{equation}
By writing this, we are done. The equations \eqref{XfromU} rewritten with the simplified potential \eqref{SimplePots2} give the solution we are looking for.
\begin{equation} \label{VTSolution}
2a \chi_{_\textsc{R}} = m (k_{_\textsc{F0}} + \textbf{k}_{_\textsc{F}}\textbf{.}\boldsymbol{\sigma}) U  
+ [\ k_{_\textsc{P0}} + \textbf{k}_{_\textsc{P}}\textbf{.}\boldsymbol{\sigma} 
- e \Phi(\textbf{n}_{_\textsc{E}} + \textbf{n}_{_\textsc{B}})\textbf{.}\boldsymbol{\sigma}\ ](k_{_\textsc{F0}} - \textbf{k}_{_\textsc{F}}\textbf{.}\boldsymbol{\sigma}) U
\end{equation}
\begin{equation*}
2a \chi_{_\textsc{L}} = m (k_{_\textsc{F0}} - \textbf{k}_{_\textsc{F}}\textbf{.}\boldsymbol{\sigma}) U  
+ [\ k_{_\textsc{P0}} - \textbf{k}_{_\textsc{P}}\textbf{.}\boldsymbol{\sigma} 
+ e \Phi(\textbf{n}_{_\textsc{E}} - \textbf{n}_{_\textsc{B}})\textbf{.}\boldsymbol{\sigma}\ ](k_{_\textsc{F0}} + \textbf{k}_{_\textsc{F}}\textbf{.}\boldsymbol{\sigma}) U
\end{equation*}

As could be expected, there is a correspondence between the solutions \eqref{VTSolution} and the solutions of the Dirac equation for a free particle.
Let $u_{_\textsc{R}}$ and $u_{_\textsc{L}}$ be the spinors $\chi_{_\textsc{R}}$ and $\chi_{_\textsc{L}}$ in the absence of an electromagnetic field
\begin{equation*}
u_{_\textsc{R}} = \chi_{_\textsc{R}}|_{\Phi = 0} \hspace{8 mm} 
u_{_\textsc{L}} = \chi_{_\textsc{L}}|_{\Phi = 0}
\end{equation*}
When $\Phi = 0$ \eqref{VTSolution} reduce to
\begin{equation*}
2a u_{_\textsc{R}} = m (k_{_\textsc{F0}} + \textbf{k}_{_\textsc{F}}\textbf{.}\boldsymbol{\sigma}) U_0 
+ ( k_{_\textsc{P0}} + \textbf{k}_{_\textsc{P}}\textbf{.}\boldsymbol{\sigma} )
(k_{_\textsc{F0}} - \textbf{k}_{_\textsc{F}}\textbf{.}\boldsymbol{\sigma}) U_0
\end{equation*}
\begin{equation*}
2a u_{_\textsc{L}} = m (k_{_\textsc{F0}} - \textbf{k}_{_\textsc{F}}\textbf{.}\boldsymbol{\sigma}) U_0 
+ ( k_{_\textsc{P0}} - \textbf{k}_{_\textsc{P}}\textbf{.}\boldsymbol{\sigma} )
(k_{_\textsc{F0}} + \textbf{k}_{_\textsc{F}}\textbf{.}\boldsymbol{\sigma}) U_0
\end{equation*}
It follows that
\begin{equation} \label{AmplitudeDiracEquation}
( k_{_\textsc{P0}} - \textbf{k}_{_\textsc{P}}\textbf{.}\boldsymbol{\sigma} )u_{_\textsc{R}} = m u_{_\textsc{L}}
\end{equation}
\begin{equation*}
( k_{_\textsc{P0}} + \textbf{k}_{_\textsc{P}}\textbf{.}\boldsymbol{\sigma} )u_{_\textsc{L}} = m u_{_\textsc{R}}
\end{equation*}
Therefore, the spinors
\begin{equation}
\psi_{_\textsc{R}} = 
u_{_\textsc{R}} e^{i(\textbf{k}_{_\textsc{P}}\textbf{.r}\ - \omega_{_\textsc{P}} t)}
\hspace{10 mm}
\psi_{_\textsc{L}} = 
u_{_\textsc{L}} e^{i(\textbf{k}_{_\textsc{P}}\textbf{.r}\ - \omega_{_\textsc{P}} t)}
\end{equation}
are a solution of the Dirac equation for a free particle. We can express the arbitrary constant spinor $U_0$ by the amplitudes $u_{_\textsc{R}}$ and $u_{_\textsc{L}}$ from the corresponding wave function of a free particle.
\begin{equation} \label{U0byAmplitudes}
2U_0 = (1 - \textbf{n.}\boldsymbol{\sigma})u_{_\textsc{R}} 
+ (1 + \textbf{n.}\boldsymbol{\sigma})u_{_\textsc{L}}
\end{equation}

Instead of expressing the results by arbitrary constants we will use the tensors (bilinear covariants) of a free particle to parametrize these results. 
Perhaps this will make easier to realize what does the electromagnetic wave does with the observables of a free Dirac particle. We use the scalar $s$, Dirac current $J_\textsc{D}$ and axial current $J_\textsc{A}$ of a free particle. These are defined in the second section.

We actually need the expressions $U_0^\dagger U_0$ and $U_0^\dagger\boldsymbol{\sigma} U_0$. From \eqref{U0byAmplitudes} we obtain.
\begin{equation*}
\begin{split}
& 2U_0^\dagger U_0 
= u_{_\textsc{R}}^\dagger u_{_\textsc{R}} + u_{_\textsc{L}}^\dagger u_{_\textsc{L}} -
\textbf{n.}(u_{_\textsc{R}}^\dagger \boldsymbol{\sigma} u_{_\textsc{R}} - u_{_\textsc{L}}^\dagger \boldsymbol{\sigma} u_{_\textsc{L}}) \\
& 2U_0^\dagger\boldsymbol{\sigma} U_0 = 
- [\ u_{_\textsc{R}}^\dagger u_{_\textsc{R}} - u_{_\textsc{L}}^\dagger u_{_\textsc{L}}
- \textbf{n.}(u_{_\textsc{R}}^\dagger \boldsymbol{\sigma} u_{_\textsc{R}} + u_{_\textsc{L}}^\dagger \boldsymbol{\sigma} u_{_\textsc{L}})\ ]\textbf{n} \\
& \hspace{18 mm} - i\textbf{n} \times [\ u_{_\textsc{L}}^\dagger \boldsymbol{\sigma} u_{_\textsc{R}} - u_{_\textsc{R}}^\dagger \boldsymbol{\sigma} u_{_\textsc{L}} - i\textbf{n} \times 
(u_{_\textsc{L}}^\dagger \boldsymbol{\sigma} u_{_\textsc{R}} + u_{_\textsc{R}}^\dagger\boldsymbol{\sigma} u_{_\textsc{L}})\ ]
\end{split}
\end{equation*}
The spinors $u_{_\textsc{R}}$ and $u_{_\textsc{L}}$ satisfy amplitude Dirac equation \eqref{AmplitudeDiracEquation}. It follows that the Dirac current is proportional to the wave vector (or impulse $p = \hbar k_{_\textsc{P}}$).
\begin{equation*}
J_\textsc{D}^{\mu} = \frac{2s}{m} k_{_\textsc{P}}^{\mu}
= \frac{2s}{m_0 c} p^{\mu}
\hspace{10 mm}
p = (p_0, \textbf{p}) 
= (\frac{\varepsilon}{c}, \textbf{p}) = \hbar(\frac{\omega_{_\textsc{P}}}{c}, \textbf{k}_{_\textsc{P}})
\end{equation*}
The scalar $s$ depends on the normalization. If the normalization condition is
$J_{D0} = 1$ then
\begin{equation*}
s = \frac{m_0c^2}{2\hbar \omega_{_\textsc{P}}} = \frac{m_0c^2}{2\varepsilon}
\end{equation*}
It also follows from \eqref{AmplitudeDiracEquation} that
\begin{equation*}
u_{_\textsc{L}}^\dagger \boldsymbol{\sigma} u_{_\textsc{R}} = 
(u_{_\textsc{R}}^\dagger \boldsymbol{\sigma} u_{_\textsc{L}})^* = 
\frac{1}{2m}(k_{_\textsc{P0}} \textbf{J}_\textsc{A} - J_\textsc{A0} \textbf{k}_{_\textsc{P}} 
+ i \textbf{k}_{_\textsc{P}} \times \textbf{J}_\textsc{A})
\end{equation*}
Finally we arrive at
\begin{equation} \label{U0byBilinears}
2U_0^\dagger U_0 = J_\textsc{D0} - \textbf{n.J}_\textsc{D}
\end{equation}
\begin{equation*}
2U_0^\dagger\boldsymbol{\sigma} U_0 = -(J_\textsc{A0} - \textbf{n.J}_\textsc{A}) \textbf{n}
+ \frac{1}{2s} \textbf{n} \times [\ \textbf{J}_\textsc{D} \times \textbf{J}_\textsc{A} 
+ \textbf{n} \times (J_\textsc{A0}\textbf{J}_\textsc{D} - J_\textsc{D0}\textbf{J}_\textsc{A})\ ]
\end{equation*}

We can now directly compare our solution to the conventional one. This gives an idea of what to expect from the modification of the coupling term in the Dirac equation. In what follows, primes denote the quantities for the particle in the electromagnetic wave.

\vspace{2 mm}
\textbf{Volkov's probability current}
\vspace{1 mm}

When solving the conventional problem, one has to put $A_\textsc{B} = 0$ 
in the general expressions \eqref{ProbabilityCurrent} and \eqref{DiffEqForV}. 
The potential $A_\textsc{E}$ have to be doubled to set the same electromagnetic field and to get directly comparable results. The right-hand side of the differential equation \eqref{DiffEqForV} is zero and the spinor $U$ is $U = U_0 e^{iS}$.
The probability current \eqref{ProbabilityCurrent} simplifies to
\begin{equation*}
(k_{_\textsc{P0}} - \textbf{n.k}_{_\textsc{P}})^2 J_\textsc{D0}^{\prime} = 
(k_{_\textsc{P0}} - \textbf{n.k}_{_\textsc{P}}) k_{_\textsc{P0}} 2U_0^\dagger U_0 + 
2e \Phi (e \Phi - \textbf{k}_{_\textsc{P}} \textbf{.n}_{_\textsc{E}} ) 2U_0^\dagger U_0
\end{equation*}
\begin{equation*}
(k_{_\textsc{P0}} - \textbf{n.k}_{_\textsc{P}})^2 \textbf{J}_\textsc{D}^{\prime} = 
(k_{_\textsc{P0}} - \textbf{n.k}_{_\textsc{P}}) (\textbf{k}_{_\textsc{P}} - 2e \Phi \textbf{n}_{_\textsc{E}}) 2U_0^\dagger U_0 + 
2e \Phi (e \Phi - \textbf{k}_{_\textsc{P}} \textbf{.n}_{_\textsc{E}} ) 2U_0^\dagger U_0 \textbf{n}
\end{equation*}
Using \eqref{U0byBilinears} this can be written as
\begin{equation} \label{VolkovCurrent}
J_\textsc{D0}^\prime = J_\textsc{D0} + 
2V \frac{V J_\textsc{D0} - \textbf{n}_{_\textsc{E}}\textbf{.J}_\textsc{D}}{J_\textsc{D0} - \textbf{n.J}_\textsc{D}}J_\textsc{D0}
\end{equation}
\begin{equation*}
\textbf{J}_\textsc{D}^\prime = \textbf{J}_\textsc{D}
+ 2V \frac{V J_\textsc{D0} - \textbf{n}_{_\textsc{E}}\textbf{.J}_\textsc{D}}{J_\textsc{D0} - \textbf{n.J}_\textsc{D}} J_\textsc{D0} \textbf{n} 
- 2V J_\textsc{D0} \textbf{n}_{_\textsc{E}}
\end{equation*}
where we have put
\begin{equation*}
V = \frac{q\Phi}{\hbar \omega_{_\textsc{P}}} c
\end{equation*}

It is seen that Volkov current is determined only by the Dirac current of the free particle. Other details of the free wave function do not matter.
Two free wave functions with equal impulses and different spins will result in the same probability current \eqref{VolkovCurrent}. So Volkov current is \emph{spin independent}.

Let us specify function $\Phi$ as a simple periodic function.
\begin{equation} \label{SimplePhi}
\Phi(\zeta) = \frac{E_{max}}{2\omega_{_\textsc{F}}} \cos(\zeta)
\end{equation}
where $E_{max}$ is the maximum electric field of the electromagnetic wave.
In this case, all terms in \eqref{VolkovCurrent}, which are linear with respect to $V$, oscillate following the oscillations of the electromagnetic field and do not generate significant current through a fixed surface element. The only significant term is the one that contains $V^2$.
To eliminate the oscillating terms, one can average the current over a period of the electromagnetic wave $T$.
\begin{equation*}
\overline{\textbf{J}_\textsc{D}^{\prime}} =
\frac{1}{T} \int_{0}^{T} dt \hspace{0.8 mm} \textbf{J}_\textsc{D}^{\prime} =
\textbf{J}_\textsc{D} +
\left( \frac{q E_{max}}{2 \omega_{_\textsc{F}}} \right)^2 
\frac{J_\textsc{D0}}{p_0(p_0 - \textbf{n.p})} \textbf{n}
\end{equation*}

\vspace{2 mm}
\textbf{Probability current with the modified coupling term}
\vspace{1 mm}

From the general expression for the probability current \eqref{ProbabilityCurrent},
inserting the simplified potentials \eqref{SimplePots2} we obtain
\begin{equation} \label{ProbCur1}
(k_{_\textsc{P0}} - \textbf{n.k}_{_\textsc{P}})^2 J_\textsc{D0}^\prime =
2(k_{_\textsc{P0}} - \textbf{n.k}_{_\textsc{P}}) U^\dagger U
\end{equation}
\begin{equation*}
\hspace{5 mm} 
+ \hspace{0.8 mm} 2e\Phi (e\Phi - \textbf{k}_{_\textsc{P}}\textbf{.n}_{_\textsc{E}}) U^\dagger U
- 2e\Phi (m \textbf{n}_{_\textsc{B}} - \textbf{k}_{_\textsc{P}}.\textbf{n}_{_\textsc{B}} \textbf{n} )\textbf{.}U^\dagger\boldsymbol{\sigma} U
\end{equation*}
\begin{equation*}
(k_{_\textsc{P0}} - \textbf{n.k}_{_\textsc{P}})^2 \textbf{J}_\textsc{D}^\prime =
2(k_{_\textsc{P0}} - \textbf{n.k}_{_\textsc{P}}) (\textbf{k}_{_\textsc{P}} - e\Phi\textbf{n}_{_\textsc{E}}) U^\dagger U
\end{equation*}
\begin{equation*}
\hspace{5 mm} 
+ \hspace{0.8 mm} 2e\Phi (e\Phi - \textbf{k}_{_\textsc{P}}\textbf{.n}_{_\textsc{E}}) U^\dagger U \textbf{n}
- 2e\Phi (m \textbf{n}_{_\textsc{B}} - \textbf{k}_{_\textsc{P}}.\textbf{n}_{_\textsc{B}} \textbf{n} )\textbf{.}U^\dagger\boldsymbol{\sigma} U \textbf{n}
+ 2e\Phi (k_{_\textsc{P0}} - \textbf{n.k}_{_\textsc{P}}) \textbf{n.}U^\dagger\boldsymbol{\sigma}U \textbf{n}_{_\textsc{B}} 
\end{equation*}
We need the expressions $U^\dagger U$ and $U^\dagger\boldsymbol{\sigma} U$ and we calculate them from \eqref{EquationForU}.
\begin{equation*}
U^\dagger U = U_0^\dagger U_0
\end{equation*}
\begin{equation*}
U^\dagger\boldsymbol{\sigma} U = U_0^\dagger \boldsymbol{\sigma} U_0
\end{equation*}
\begin{equation*}
\hspace{5 mm}
+ \hspace{0.8 mm} \frac{2}{\beta^2} \sin(\beta\eta) (m \textbf{n}_{_\textsc{B}} - \textbf{k}_{_\textsc{P}}\textbf{.n}_{_\textsc{B}} \textbf{n}) \times
[\ \beta \cos(\beta\eta)U_0^\dagger \boldsymbol{\sigma} U_0 + \sin(\beta\eta)(m \textbf{n}_{_\textsc{B}} - \textbf{k}_{_\textsc{P}}\textbf{.n}_{_\textsc{B}} \textbf{n}) \times U_0^\dagger \boldsymbol{\sigma} U_0\ ]
\end{equation*}
We replace these expressions in \eqref{ProbCur1} and obtain
\begin{equation} \label{ProbCur2}
(k_{_\textsc{P0}} - \textbf{n.k}_{_\textsc{P}})^2 J_\textsc{D0}^\prime =
2(k_{_\textsc{P0}} - \textbf{n.k}_{_\textsc{P}}) U_0^\dagger U_0
\end{equation}
\begin{equation*}
\hspace{5 mm} 
+ \hspace{0.8 mm} 2e\Phi (e\Phi - \textbf{k}_{_\textsc{P}}\textbf{.n}_{_\textsc{E}}) U_0^\dagger U_0
- 2e\Phi (m \textbf{n}_{_\textsc{B}} - \textbf{k}_{_\textsc{P}}.\textbf{n}_{_\textsc{B}} \textbf{n} )\textbf{.}U_0^\dagger\boldsymbol{\sigma} U_0
\end{equation*}
\begin{equation*}
(k_{_\textsc{P0}} - \textbf{n.k}_{_\textsc{P}})^2 \textbf{J}_\textsc{D}^\prime =
2(k_{_\textsc{P0}} - \textbf{n.k}_{_\textsc{P}}) (\textbf{k}_{_\textsc{P}} - e\Phi\textbf{n}_{_\textsc{E}}) U_0^\dagger U_0
\end{equation*}
\begin{equation*}
\hspace{5 mm} 
+ \hspace{0.8 mm} 2e\Phi (e\Phi - \textbf{k}_{_\textsc{P}}\textbf{.n}_{_\textsc{E}}) U_0^\dagger U_0 \textbf{n}
- 2e\Phi (m \textbf{n}_{_\textsc{B}} - \textbf{k}_{_\textsc{P}}.\textbf{n}_{_\textsc{B}} \textbf{n} )\textbf{.}U_0^\dagger\boldsymbol{\sigma} U_0 \textbf{n}
+ 2e\Phi (k_{_\textsc{P0}} - \textbf{n.k}_{_\textsc{P}}) \textbf{n.}U_0^\dagger\boldsymbol{\sigma}U_0 \textbf{n}_{_\textsc{B}} 
\end{equation*}
\begin{equation*}
\hspace{5 mm} 
- \hspace{0.8 mm} \frac{4 m e \Phi}{\beta^2} (k_{_\textsc{P0}} - \textbf{n.k}_{_\textsc{P}}) \sin(\beta\eta)
[\ \sin(\beta\eta)(m \textbf{n}_{_\textsc{B}} - \textbf{k}_{_\textsc{P}}.\textbf{n}_{_\textsc{B}} \textbf{n} ) + \beta \cos(\beta\eta)\textbf{n}_{_\textsc{E}}\ ]\textbf{.} U_0^\dagger\boldsymbol{\sigma}U_0 \textbf{n}_{_\textsc{B}} 
\end{equation*}
And using \eqref{U0byBilinears} we arrive at
\begin{equation} \label{MyCurrent}
J_\textsc{D0}^\prime = J_\textsc{D0} + 
V \frac{J_\textsc{D0} V - \textbf{n}_{_\textsc{E}}\textbf{.J}_\textsc{D} 
- \textbf{n}_{_\textsc{B}}\textbf{.J}_\textsc{A}}{J_\textsc{D0} - \textbf{n.J}_\textsc{D}}J_\textsc{D0}
\end{equation}
\begin{equation*}
\textbf{J}_\textsc{D}^\prime = \textbf{J}_\textsc{D} + 
V \frac{J_\textsc{D0} V - \textbf{n}_{_\textsc{E}} \textbf{.J}_\textsc{D} 
- \textbf{n}_{_\textsc{B}}\textbf{.J}_\textsc{A}}{J_\textsc{D0} - \textbf{n.J}_\textsc{D}} J_\textsc{D0} \textbf{n}
- V J_{\textsc{D0}} \textbf{n}_{_\textsc{E}} 
- V \frac{J_\textsc{A0} - \textbf{n.J}_\textsc{A}}{J_\textsc{D0} - \textbf{n.J}_\textsc{D}} J_{\textsc{D0}} \textbf{n}_{_\textsc{B}}
\end{equation*}
\begin{equation*}
\hspace{5 mm}
- \hspace{0.8 mm} 2m \frac{V \omega_{_\textsc{P}}}{c \beta^2} \sin^2(\beta\eta)
\textbf{n}_{_\textsc{B}}\textbf{.J}_\textsc{A} \textbf{n}_{_\textsc{B}}
+ \frac{V \omega_{_\textsc{P}}}{c \beta} \sin(2\beta\eta)
\frac { (J_\textsc{A0} - \textbf{n.J}_\textsc{A}) \textbf{n}_{_\textsc{E}}\textbf{.J}_\textsc{D} - (J_\textsc{D0} - \textbf{n.J}_\textsc{D}) \textbf{n}_{_\textsc{E}}\textbf{.J}_\textsc{A}}{J_\textsc{D0} - \textbf{n.J}_\textsc{D}}
\textbf{n}_{_\textsc{B}}
\end{equation*}

It is seen that the probability current is determined by both Dirac current and axial current of the free particle. The details of the free wave function matter. Two free wave functions with equal impulses and different spins will result in different probability currents \eqref{MyCurrent}. So, in general, the probability current is \emph{spin dependent}.

Let $\Phi$ is defined by \eqref{SimplePhi}. Averaging the probability current over one period of the electromagnetic wave eliminates all terms, which are linear with respect to V, including those containing $\sin^2(\beta\eta)$ and $\sin(2\beta\eta)$. The result is.
\begin{equation*}
\overline{\textbf{J}_\textsc{D}^{\prime}} =
\textbf{J}_\textsc{D} +
\frac{1}{2} \left( \frac{q E_{max}}{2 \omega_{_\textsc{F}}} \right)^2 
\frac{J_\textsc{D0}}{p_0(p_0 - \textbf{n.p})} \textbf{n}
\end{equation*}
The average component of the probability current in the direction of the Poynting vector is two times smaller than the corresponding component of the Volkov's current.

\section{Acknowledgements}
I would like to thank Prof. Georgi Raikov for the valuable advice and constructive discussion. Without his help this work would not have been completed and published.

\end{document}